\documentclass[%
 reprint,
 superscriptaddress,
nofootinbib,
 amsmath,amssymb,
 aps,
 normalem
]{revtex4-1}

\usepackage{graphicx}
\usepackage{dcolumn}
\usepackage{bm}
\usepackage{hhline}
\usepackage{xcolor}
\usepackage{xstring}
\usepackage{xparse}
\usepackage{siunitx}
\usepackage{hyperref}
\usepackage{longtable}
\usepackage{tabularx}
\usepackage{orcidlink}

\newcommand{\chieff}{\chi_{\mathrm{eff}}}
\newcommand{\chip}{\chi_\mathrm{p}}

\ExplSyntaxOn
\NewDocumentCommand{\longdash}{ O{2} }
 {
  --\prg_replicate:nn { #1 - 1 } { \negthinspace -- }
 }
\ExplSyntaxOff

\newcounter{RunIDCounter}

\newcommand{\ridref}[1]{\refstepcounter{RunIDCounter}\label{#1}\theRunIDCounter}
\definecolor{shgreen}{rgb}{0.15625, 0.609375, 0.316406}

\newcommand{\bw}[0]{\textsc{bayeswave}}
\newcommand{\bilby}[0]{\textsc{bilby}}
\newcommand{\hmmslong}[0]{\textsc{HighMass-ModerateSpins}}
\newcommand{\hmeslong}[0]{\textsc{HighMass-ExtremeSpins}}
\newcommand{\mmmslong}[0]{\textsc{ModerateMass-ModerateSpins}}
\newcommand{\mmeslong}[0]{\textsc{ModerateMass-ExtremeSpins}}
\newcommand{\hmms}[0]{\textsc{HMMS}}
\newcommand{\hmes}[0]{\textsc{HMES}}
\newcommand{\mmms}[0]{\textsc{MMMS}}
\newcommand{\mmes}[0]{\textsc{MMES}}

\newcommand{\runref}[1]{(Run~\ref{#1} in Table~\ref{tab:table-of-runs})}
\newcommand{\runsref}[2]{(Runs~\ref{#1}---\ref{#2} in Table~\ref{tab:table-of-runs})}

\newcommand{\testifthen}[1]{\IfEqCase{#1}{{a}{1}{b}{2}}}

\DeclareRobustCommand{\PTCHIEFF}[1]{\IfEqCase{#1}{{ZHMMSNW}{90\%}{ZHMMST3P0}{4\%}{ZHMMST3P1}{52\%}{ZHMMST3P2}{16\%}{ZHMMST3P3}{1\%}{ZHMMST3P4}{25\%}{ZHMMST3P5}{1\%}{ZHMMST2P0}{22\%}{ZHMMST2P1}{1\%}{ZHMMST2P2}{17\%}{ZHMMST2P3}{75\%}{ZHMMST2P4}{6\%}{ZHMMST2P5}{1\%}{ZHMMST1P0}{10\%}{ZHMMST1P1}{12\%}{ZHMMST1P2}{66\%}{ZHMMST1P3}{13\%}{ZHMMST1P4}{74\%}{ZHMMST1P5}{43\%}{ZHMMSMIN}{1\%}{ZHMMSMAX}{90\%}{ZHMESNW}{7\%}{ZHMEST3P0}{0\%}{ZHMEST3P1}{0\%}{ZHMEST3P2}{6\%}{ZHMEST3P3}{61\%}{ZHMEST3P4}{0\%}{ZHMEST3P5}{0\%}{ZHMEST2P0}{97\%}{ZHMEST2P1}{0\%}{ZHMEST2P2}{0\%}{ZHMEST2P3}{0\%}{ZHMEST2P4}{15\%}{ZHMEST2P5}{75\%}{ZHMEST1P0}{23\%}{ZHMEST1P1}{98\%}{ZHMEST1P2}{87\%}{ZHMEST1P3}{0\%}{ZHMEST1P4}{0\%}{ZHMEST1P5}{0\%}{ZHMESMIN}{0\%}{ZHMESMAX}{98\%}{ZMMMSNW}{93\%}{ZMMMST3P0}{75\%}{ZMMMST3P1}{6\%}{ZMMMST3P2}{1\%}{ZMMMST3P3}{40\%}{ZMMMST3P4}{11\%}{ZMMMST3P5}{11\%}{ZMMMST2P0}{12\%}{ZMMMST2P1}{40\%}{ZMMMST2P2}{1\%}{ZMMMST2P3}{1\%}{ZMMMST2P4}{42\%}{ZMMMST2P5}{1\%}{ZMMMST1P0}{3\%}{ZMMMST1P1}{34\%}{ZMMMST1P2}{21\%}{ZMMMST1P3}{1\%}{ZMMMST1P4}{13\%}{ZMMMST1P5}{23\%}{ZMMMSMIN}{1\%}{ZMMMSMAX}{93\%}{ZMMESNW}{21\%}{ZMMEST3P0}{0\%}{ZMMEST3P1}{99\%}{ZMMEST3P2}{10\%}{ZMMEST3P3}{41\%}{ZMMEST3P4}{4\%}{ZMMEST3P5}{0\%}{ZMMEST2P0}{0\%}{ZMMEST2P1}{5\%}{ZMMEST2P2}{60\%}{ZMMEST2P3}{4\%}{ZMMEST2P4}{72\%}{ZMMEST2P5}{0\%}{ZMMEST1P0}{0\%}{ZMMEST1P1}{0\%}{ZMMEST1P2}{6\%}{ZMMEST1P3}{71\%}{ZMMEST1P4}{8\%}{ZMMEST1P5}{53\%}{ZMMESMIN}{0\%}{ZMMESMAX}{99\%}{GHMMSNW}{33\%}{GHMMST3P0}{4\%}{GHMMST3P1}{60\%}{GHMMST3P2}{2\%}{GHMMST3P3}{0\%}{GHMMST3P4}{54\%}{GHMMST3P5}{10\%}{GHMMST2P0}{33\%}{GHMMST2P1}{15\%}{GHMMST2P2}{22\%}{GHMMST2P3}{68\%}{GHMMST2P4}{0\%}{GHMMST2P5}{0\%}{GHMMST1P0}{0\%}{GHMMST1P1}{0\%}{GHMMST1P2}{81\%}{GHMMST1P3}{72\%}{GHMMST1P4}{82\%}{GHMMST1P5}{10\%}{GHMMSMIN}{0\%}{GHMMSMAX}{82\%}{GMMMSNW}{6\%}{GMMMST3P0}{18\%}{GMMMST3P1}{1\%}{GMMMST3P2}{0\%}{GMMMST3P3}{3\%}{GMMMST3P4}{97\%}{GMMMST3P5}{92\%}{GMMMST2P0}{76\%}{GMMMST2P1}{2\%}{GMMMST2P2}{0\%}{GMMMST2P3}{0\%}{GMMMST2P4}{16\%}{GMMMST2P5}{40\%}{GMMMST1P0}{91\%}{GMMMST1P1}{37\%}{GMMMST1P2}{1\%}{GMMMST1P3}{0\%}{GMMMST1P4}{0\%}{GMMMST1P5}{34\%}{GMMMSMIN}{0\%}{GMMMSMAX}{97\%}}[\textcolor{red}{???}]}
\DeclareRobustCommand{\PTCHIP}[1]{\IfEqCase{#1}{{ZHMMSNW}{97\%}{ZHMMST3P0}{22\%}{ZHMMST3P1}{10\%}{ZHMMST3P2}{53\%}{ZHMMST3P3}{68\%}{ZHMMST3P4}{3\%}{ZHMMST3P5}{82\%}{ZHMMST2P0}{9\%}{ZHMMST2P1}{88\%}{ZHMMST2P2}{66\%}{ZHMMST2P3}{54\%}{ZHMMST2P4}{97\%}{ZHMMST2P5}{33\%}{ZHMMST1P0}{88\%}{ZHMMST1P1}{23\%}{ZHMMST1P2}{31\%}{ZHMMST1P3}{85\%}{ZHMMST1P4}{79\%}{ZHMMST1P5}{98\%}{ZHMMSMIN}{3\%}{ZHMMSMAX}{98\%}{ZHMESNW}{85\%}{ZHMEST3P0}{59\%}{ZHMEST3P1}{32\%}{ZHMEST3P2}{61\%}{ZHMEST3P3}{63\%}{ZHMEST3P4}{6\%}{ZHMEST3P5}{89\%}{ZHMEST2P0}{23\%}{ZHMEST2P1}{90\%}{ZHMEST2P2}{71\%}{ZHMEST2P3}{93\%}{ZHMEST2P4}{86\%}{ZHMEST2P5}{74\%}{ZHMEST1P0}{93\%}{ZHMEST1P1}{87\%}{ZHMEST1P2}{46\%}{ZHMEST1P3}{54\%}{ZHMEST1P4}{47\%}{ZHMEST1P5}{71\%}{ZHMESMIN}{6\%}{ZHMESMAX}{93\%}{ZMMMSNW}{80\%}{ZMMMST3P0}{3\%}{ZMMMST3P1}{68\%}{ZMMMST3P2}{86\%}{ZMMMST3P3}{1\%}{ZMMMST3P4}{4\%}{ZMMMST3P5}{6\%}{ZMMMST2P0}{25\%}{ZMMMST2P1}{18\%}{ZMMMST2P2}{45\%}{ZMMMST2P3}{41\%}{ZMMMST2P4}{1\%}{ZMMMST2P5}{42\%}{ZMMMST1P0}{60\%}{ZMMMST1P1}{53\%}{ZMMMST1P2}{85\%}{ZMMMST1P3}{51\%}{ZMMMST1P4}{38\%}{ZMMMST1P5}{7\%}{ZMMMSMIN}{1\%}{ZMMMSMAX}{86\%}{ZMMESNW}{7\%}{ZMMEST3P0}{48\%}{ZMMEST3P1}{48\%}{ZMMEST3P2}{17\%}{ZMMEST3P3}{43\%}{ZMMEST3P4}{8\%}{ZMMEST3P5}{2\%}{ZMMEST2P0}{0\%}{ZMMEST2P1}{27\%}{ZMMEST2P2}{18\%}{ZMMEST2P3}{10\%}{ZMMEST2P4}{48\%}{ZMMEST2P5}{11\%}{ZMMEST1P0}{11\%}{ZMMEST1P1}{0\%}{ZMMEST1P2}{2\%}{ZMMEST1P3}{3\%}{ZMMEST1P4}{19\%}{ZMMEST1P5}{53\%}{ZMMESMIN}{0\%}{ZMMESMAX}{53\%}{GHMMSNW}{67\%}{GHMMST3P0}{13\%}{GHMMST3P1}{7\%}{GHMMST3P2}{44\%}{GHMMST3P3}{59\%}{GHMMST3P4}{19\%}{GHMMST3P5}{86\%}{GHMMST2P0}{50\%}{GHMMST2P1}{99\%}{GHMMST2P2}{38\%}{GHMMST2P3}{29\%}{GHMMST2P4}{54\%}{GHMMST2P5}{39\%}{GHMMST1P0}{28\%}{GHMMST1P1}{27\%}{GHMMST1P2}{79\%}{GHMMST1P3}{51\%}{GHMMST1P4}{82\%}{GHMMST1P5}{64\%}{GHMMSMIN}{7\%}{GHMMSMAX}{99\%}{GMMMSNW}{69\%}{GMMMST3P0}{3\%}{GMMMST3P1}{3\%}{GMMMST3P2}{11\%}{GMMMST3P3}{1\%}{GMMMST3P4}{3\%}{GMMMST3P5}{47\%}{GMMMST2P0}{76\%}{GMMMST2P1}{11\%}{GMMMST2P2}{48\%}{GMMMST2P3}{4\%}{GMMMST2P4}{1\%}{GMMMST2P5}{17\%}{GMMMST1P0}{30\%}{GMMMST1P1}{85\%}{GMMMST1P2}{78\%}{GMMMST1P3}{70\%}{GMMMST1P4}{5\%}{GMMMST1P5}{1\%}{GMMMSMIN}{1\%}{GMMMSMAX}{85\%}}[\textcolor{red}{???}]}
\DeclareRobustCommand{\PTCONLABELS}[1]{\IfEqCase{#1}{{ZHMMST3P0}{$t=t_3,\phi=0$}{ZHMMST3P1}{$t=t_3,\phi=\frac{\pi}{3}$}{ZHMMST3P2}{$t=t_3,\phi=\frac{2\pi}{3}$}{ZHMMST3P3}{$t=t_3,\phi=\pi$}{ZHMMST3P4}{$t=t_3,\phi=\frac{4\pi}{3}$}{ZHMMST3P5}{$t=t_3,\phi=\frac{5\pi}{3}$}{ZHMMST2P0}{$t=t_2,\phi=0$}{ZHMMST2P1}{$t=t_2,\phi=\frac{\pi}{3}$}{ZHMMST2P2}{$t=t_2,\phi=\frac{2\pi}{3}$}{ZHMMST2P3}{$t=t_2,\phi=\pi$}{ZHMMST2P4}{$t=t_2,\phi=\frac{4\pi}{3}$}{ZHMMST2P5}{$t=t_2,\phi=\frac{5\pi}{3}$}{ZHMMST1P0}{$t=t_1,\phi=0$}{ZHMMST1P1}{$t=t_1,\phi=\frac{\pi}{3}$}{ZHMMST1P2}{$t=t_1,\phi=\frac{2\pi}{3}$}{ZHMMST1P3}{$t=t_1,\phi=\pi$}{ZHMMST1P4}{$t=t_1,\phi=\frac{4\pi}{3}$}{ZHMMST1P5}{$t=t_1,\phi=\frac{5\pi}{3}$}{ZHMMSMINCHIEFF}{$t=t_3,\phi=\frac{5\pi}{3}$}{ZHMMSMINCHIP}{$t=t_3,\phi=\frac{4\pi}{3}$}{ZHMMSMAXCHIEFF}{CBC only}{ZHMMSMAXCHIP}{$t=t_1,\phi=\frac{5\pi}{3}$}{ZHMEST3P0}{$t=t_3,\phi=0$}{ZHMEST3P1}{$t=t_3,\phi=\frac{\pi}{3}$}{ZHMEST3P2}{$t=t_3,\phi=\frac{2\pi}{3}$}{ZHMEST3P3}{$t=t_3,\phi=\pi$}{ZHMEST3P4}{$t=t_3,\phi=\frac{4\pi}{3}$}{ZHMEST3P5}{$t=t_3,\phi=\frac{5\pi}{3}$}{ZHMEST2P0}{$t=t_2,\phi=0$}{ZHMEST2P1}{$t=t_2,\phi=\frac{\pi}{3}$}{ZHMEST2P2}{$t=t_2,\phi=\frac{2\pi}{3}$}{ZHMEST2P3}{$t=t_2,\phi=\pi$}{ZHMEST2P4}{$t=t_2,\phi=\frac{4\pi}{3}$}{ZHMEST2P5}{$t=t_2,\phi=\frac{5\pi}{3}$}{ZHMEST1P0}{$t=t_1,\phi=0$}{ZHMEST1P1}{$t=t_1,\phi=\frac{\pi}{3}$}{ZHMEST1P2}{$t=t_1,\phi=\frac{2\pi}{3}$}{ZHMEST1P3}{$t=t_1,\phi=\pi$}{ZHMEST1P4}{$t=t_1,\phi=\frac{4\pi}{3}$}{ZHMEST1P5}{$t=t_1,\phi=\frac{5\pi}{3}$}{ZHMESMINCHIEFF}{$t=t_2,\phi=\frac{2\pi}{3}$}{ZHMESMINCHIP}{$t=t_3,\phi=\frac{4\pi}{3}$}{ZHMESMAXCHIEFF}{$t=t_1,\phi=\frac{\pi}{3}$}{ZHMESMAXCHIP}{$t=t_2,\phi=\pi$}{ZMMMST3P0}{$t=t_3,\phi=0$}{ZMMMST3P1}{$t=t_3,\phi=\frac{\pi}{3}$}{ZMMMST3P2}{$t=t_3,\phi=\frac{2\pi}{3}$}{ZMMMST3P3}{$t=t_3,\phi=\pi$}{ZMMMST3P4}{$t=t_3,\phi=\frac{4\pi}{3}$}{ZMMMST3P5}{$t=t_3,\phi=\frac{5\pi}{3}$}{ZMMMST2P0}{$t=t_2,\phi=0$}{ZMMMST2P1}{$t=t_2,\phi=\frac{\pi}{3}$}{ZMMMST2P2}{$t=t_2,\phi=\frac{2\pi}{3}$}{ZMMMST2P3}{$t=t_2,\phi=\pi$}{ZMMMST2P4}{$t=t_2,\phi=\frac{4\pi}{3}$}{ZMMMST2P5}{$t=t_2,\phi=\frac{5\pi}{3}$}{ZMMMST1P0}{$t=t_1,\phi=0$}{ZMMMST1P1}{$t=t_1,\phi=\frac{\pi}{3}$}{ZMMMST1P2}{$t=t_1,\phi=\frac{2\pi}{3}$}{ZMMMST1P3}{$t=t_1,\phi=\pi$}{ZMMMST1P4}{$t=t_1,\phi=\frac{4\pi}{3}$}{ZMMMST1P5}{$t=t_1,\phi=\frac{5\pi}{3}$}{ZMMMSMINCHIEFF}{$t=t_2,\phi=\frac{2\pi}{3}$}{ZMMMSMINCHIP}{$t=t_3,\phi=\pi$}{ZMMMSMAXCHIEFF}{CBC only}{ZMMMSMAXCHIP}{$t=t_3,\phi=\frac{2\pi}{3}$}{ZMMEST3P0}{$t=t_3,\phi=0$}{ZMMEST3P1}{$t=t_3,\phi=\frac{\pi}{3}$}{ZMMEST3P2}{$t=t_3,\phi=\frac{2\pi}{3}$}{ZMMEST3P3}{$t=t_3,\phi=\pi$}{ZMMEST3P4}{$t=t_3,\phi=\frac{4\pi}{3}$}{ZMMEST3P5}{$t=t_3,\phi=\frac{5\pi}{3}$}{ZMMEST2P0}{$t=t_2,\phi=0$}{ZMMEST2P1}{$t=t_2,\phi=\frac{\pi}{3}$}{ZMMEST2P2}{$t=t_2,\phi=\frac{2\pi}{3}$}{ZMMEST2P3}{$t=t_2,\phi=\pi$}{ZMMEST2P4}{$t=t_2,\phi=\frac{4\pi}{3}$}{ZMMEST2P5}{$t=t_2,\phi=\frac{5\pi}{3}$}{ZMMEST1P0}{$t=t_1,\phi=0$}{ZMMEST1P1}{$t=t_1,\phi=\frac{\pi}{3}$}{ZMMEST1P2}{$t=t_1,\phi=\frac{2\pi}{3}$}{ZMMEST1P3}{$t=t_1,\phi=\pi$}{ZMMEST1P4}{$t=t_1,\phi=\frac{4\pi}{3}$}{ZMMEST1P5}{$t=t_1,\phi=\frac{5\pi}{3}$}{ZMMESMINCHIEFF}{$t=t_2,\phi=0$}{ZMMESMINCHIP}{$t=t_2,\phi=0$}{ZMMESMAXCHIEFF}{$t=t_3,\phi=\frac{\pi}{3}$}{ZMMESMAXCHIP}{$t=t_1,\phi=\frac{5\pi}{3}$}{GHMMST3P0}{$t=t_3,\phi=0$}{GHMMST3P1}{$t=t_3,\phi=\frac{\pi}{3}$}{GHMMST3P2}{$t=t_3,\phi=\frac{2\pi}{3}$}{GHMMST3P3}{$t=t_3,\phi=\pi$}{GHMMST3P4}{$t=t_3,\phi=\frac{4\pi}{3}$}{GHMMST3P5}{$t=t_3,\phi=\frac{5\pi}{3}$}{GHMMST2P0}{$t=t_2,\phi=0$}{GHMMST2P1}{$t=t_2,\phi=\frac{\pi}{3}$}{GHMMST2P2}{$t=t_2,\phi=\frac{2\pi}{3}$}{GHMMST2P3}{$t=t_2,\phi=\pi$}{GHMMST2P4}{$t=t_2,\phi=\frac{4\pi}{3}$}{GHMMST2P5}{$t=t_2,\phi=\frac{5\pi}{3}$}{GHMMST1P0}{$t=t_1,\phi=0$}{GHMMST1P1}{$t=t_1,\phi=\frac{\pi}{3}$}{GHMMST1P2}{$t=t_1,\phi=\frac{2\pi}{3}$}{GHMMST1P3}{$t=t_1,\phi=\pi$}{GHMMST1P4}{$t=t_1,\phi=\frac{4\pi}{3}$}{GHMMST1P5}{$t=t_1,\phi=\frac{5\pi}{3}$}{GHMMSMINCHIEFF}{$t=t_1,\phi=0$}{GHMMSMINCHIP}{$t=t_3,\phi=\frac{\pi}{3}$}{GHMMSMAXCHIEFF}{$t=t_1,\phi=\frac{4\pi}{3}$}{GHMMSMAXCHIP}{$t=t_2,\phi=\frac{\pi}{3}$}{GMMMST3P0}{$t=t_3,\phi=0$}{GMMMST3P1}{$t=t_3,\phi=\frac{\pi}{3}$}{GMMMST3P2}{$t=t_3,\phi=\frac{2\pi}{3}$}{GMMMST3P3}{$t=t_3,\phi=\pi$}{GMMMST3P4}{$t=t_3,\phi=\frac{4\pi}{3}$}{GMMMST3P5}{$t=t_3,\phi=\frac{5\pi}{3}$}{GMMMST2P0}{$t=t_2,\phi=0$}{GMMMST2P1}{$t=t_2,\phi=\frac{\pi}{3}$}{GMMMST2P2}{$t=t_2,\phi=\frac{2\pi}{3}$}{GMMMST2P3}{$t=t_2,\phi=\pi$}{GMMMST2P4}{$t=t_2,\phi=\frac{4\pi}{3}$}{GMMMST2P5}{$t=t_2,\phi=\frac{5\pi}{3}$}{GMMMST1P0}{$t=t_1,\phi=0$}{GMMMST1P1}{$t=t_1,\phi=\frac{\pi}{3}$}{GMMMST1P2}{$t=t_1,\phi=\frac{2\pi}{3}$}{GMMMST1P3}{$t=t_1,\phi=\pi$}{GMMMST1P4}{$t=t_1,\phi=\frac{4\pi}{3}$}{GMMMST1P5}{$t=t_1,\phi=\frac{5\pi}{3}$}{GMMMSMINCHIEFF}{$t=t_3,\phi=\frac{2\pi}{3}$}{GMMMSMINCHIP}{$t=t_2,\phi=\frac{4\pi}{3}$}{GMMMSMAXCHIEFF}{$t=t_3,\phi=\frac{4\pi}{3}$}{GMMMSMAXCHIP}{$t=t_1,\phi=\frac{\pi}{3}$}}[\textcolor{red}{???}]}

\DeclareRobustCommand{\RICHIEFF}[1]{\IfEqCase{#1}{{HMMSFD1}{33\%}{HMMSFD2}{28\%}{HMMSFD3}{18\%}{HMMSFD4}{1\%}{HMMSFD5}{1\%}{HMMSML}{15\%}{HMMSMD}{10\%}{HMMSGN}{23\%}{HMMSUS}{0\%}{HMMSJI}{34\%}{HMMSBW}{30\%}{MMMSFD1}{35\%}{MMMSFD2}{63\%}{MMMSFD3}{37\%}{MMMSFD4}{19\%}{MMMSFD5}{21\%}{MMMSML}{27\%}{MMMSMD}{27\%}{MMMSGN}{4\%}{MMMSUS}{18\%}{MMMSJI}{40\%}{MMMSBW}{18\%}}[\textcolor{red}{???}]}
\DeclareRobustCommand{\RICHIP}[1]{\IfEqCase{#1}{{HMMSFD1}{44\%}{HMMSFD2}{44\%}{HMMSFD3}{98\%}{HMMSFD4}{63\%}{HMMSFD5}{80\%}{HMMSML}{61\%}{HMMSMD}{73\%}{HMMSGN}{63\%}{HMMSUS}{59\%}{HMMSJI}{89\%}{HMMSBW}{73\%}{MMMSFD1}{65\%}{MMMSFD2}{47\%}{MMMSFD3}{72\%}{MMMSFD4}{47\%}{MMMSFD5}{74\%}{MMMSML}{58\%}{MMMSMD}{85\%}{MMMSGN}{82\%}{MMMSUS}{3\%}{MMMSJI}{61\%}{MMMSBW}{31\%}}[\textcolor{red}{???}]}
\DeclareRobustCommand{\RICHIEFFJS}[1]{\IfEqCase{#1}{{HMMSFD1}{0.024 bits}{HMMSFD2}{0.015 bits}{HMMSFD3}{0.023 bits}{HMMSFD4}{0.508 bits}{HMMSFD5}{0.331 bits}{HMMSML}{0.019 bits}{HMMSMD}{0.057 bits}{HMMSGN}{0.000 bits}{HMMSUS}{0.725 bits}{HMMSJI}{0.022 bits}{HMMSBW}{0.131 bits}{MMMSFD1}{0.291 bits}{MMMSFD2}{0.488 bits}{MMMSFD3}{0.283 bits}{MMMSFD4}{0.131 bits}{MMMSFD5}{0.158 bits}{MMMSML}{0.214 bits}{MMMSMD}{0.247 bits}{MMMSGN}{0.000 bits}{MMMSUS}{0.141 bits}{MMMSJI}{0.317 bits}{MMMSBW}{0.145 bits}}[\textcolor{red}{???}]}
\DeclareRobustCommand{\RICHIPJS}[1]{\IfEqCase{#1}{{HMMSFD1}{0.016 bits}{HMMSFD2}{0.016 bits}{HMMSFD3}{0.033 bits}{HMMSFD4}{0.151 bits}{HMMSFD5}{0.032 bits}{HMMSML}{0.006 bits}{HMMSMD}{0.009 bits}{HMMSGN}{0.000 bits}{HMMSUS}{0.155 bits}{HMMSJI}{0.024 bits}{HMMSBW}{0.053 bits}{MMMSFD1}{0.036 bits}{MMMSFD2}{0.073 bits}{MMMSFD3}{0.014 bits}{MMMSFD4}{0.022 bits}{MMMSFD5}{0.030 bits}{MMMSML}{0.019 bits}{MMMSMD}{0.012 bits}{MMMSGN}{0.000 bits}{MMMSUS}{0.524 bits}{MMMSJI}{0.020 bits}{MMMSBW}{0.054 bits}}[\textcolor{red}{???}]}
\DeclareRobustCommand{\RISNR}[1]{\IfEqCase{#1}{{HMMSMD}{4.297676861528165}{HMMSML}{4.089420375100328}{HMMSFD1}{4.654819006407808}{HMMSFD2}{4.29293888029754}{HMMSFD3}{5.950327465580117}{HMMSFD4}{6.292015224063723}{HMMSFD5}{4.5792313484556875}{HMMSUS}{7.507397626934975}{MMMSMD}{3.3744996270562653}{MMMSML}{3.496729631356438}{MMMSFD1}{3.8588467290149104}{MMMSFD2}{5.6738993437361955}{MMMSFD3}{4.448552824264665}{MMMSFD4}{3.4923970805507616}{MMMSFD5}{3.538086272319175}{MMMSUS}{7.958759755867739}}[\textcolor{red}{???}]}

\DeclareRobustCommand{\VSNRCHIEFF}[1]{\IfEqCase{#1}{{HMMSNW}{90\%}{HMMSSNR1}{74\%}{HMMSSNR2}{62\%}{HMMSSNR3}{48\%}{HMMSSNR4}{36\%}{HMMSSNR5}{20\%}{HMMSSNR6}{9\%}{HMMSSNR7}{2\%}{HMMSSNR8}{1\%}{HMMSSNR9}{0\%}{HMMSSNR10}{0\%}{HMESNW}{7\%}{HMESSNR1}{4\%}{HMESSNR2}{2\%}{HMESSNR3}{0\%}{HMESSNR4}{0\%}{HMESSNR5}{0\%}{HMESSNR6}{0\%}{HMESSNR7}{0\%}{HMESSNR8}{0\%}{HMESSNR9}{0\%}{HMESSNR10}{0\%}{MMMSNW}{93\%}{MMMSSNR1}{90\%}{MMMSSNR2}{97\%}{MMMSSNR3}{75\%}{MMMSSNR4}{88\%}{MMMSSNR5}{67\%}{MMMSSNR6}{79\%}{MMMSSNR7}{55\%}{MMMSSNR8}{35\%}{MMMSSNR9}{29\%}{MMMSSNR10}{12\%}{MMESNW}{21\%}{MMESSNR1}{8\%}{MMESSNR2}{4\%}{MMESSNR3}{2\%}{MMESSNR4}{1\%}{MMESSNR5}{0\%}{MMESSNR6}{0\%}{MMESSNR7}{0\%}{MMESSNR8}{0\%}{MMESSNR9}{0\%}{MMESSNR10}{0\%}}[\textcolor{red}{???}]}
\DeclareRobustCommand{\VSNRCHIP}[1]{\IfEqCase{#1}{{HMMSNW}{97\%}{HMMSSNR1}{84\%}{HMMSSNR2}{93\%}{HMMSSNR3}{90\%}{HMMSSNR4}{92\%}{HMMSSNR5}{52\%}{HMMSSNR6}{39\%}{HMMSSNR7}{31\%}{HMMSSNR8}{34\%}{HMMSSNR9}{34\%}{HMMSSNR10}{39\%}{HMESNW}{85\%}{HMESSNR1}{49\%}{HMESSNR2}{49\%}{HMESSNR3}{49\%}{HMESSNR4}{68\%}{HMESSNR5}{80\%}{HMESSNR6}{88\%}{HMESSNR7}{99\%}{HMESSNR8}{53\%}{HMESSNR9}{37\%}{HMESSNR10}{29\%}{MMMSNW}{80\%}{MMMSSNR1}{93\%}{MMMSSNR2}{88\%}{MMMSSNR3}{37\%}{MMMSSNR4}{31\%}{MMMSSNR5}{12\%}{MMMSSNR6}{3\%}{MMMSSNR7}{0\%}{MMMSSNR8}{1\%}{MMMSSNR9}{0\%}{MMMSSNR10}{0\%}{MMESNW}{7\%}{MMESSNR1}{3\%}{MMESSNR2}{2\%}{MMESSNR3}{3\%}{MMESSNR4}{1\%}{MMESSNR5}{1\%}{MMESSNR6}{2\%}{MMESSNR7}{2\%}{MMESSNR8}{4\%}{MMESSNR9}{9\%}{MMESSNR10}{9\%}}[\textcolor{red}{???}]}
\DeclareRobustCommand{\VSNRCHIEFFJS}[1]{\IfEqCase{#1}{{HMMSNW}{0.000 bits}{HMMSSNR1}{0.018 bits}{HMMSSNR2}{0.066 bits}{HMMSSNR3}{0.120 bits}{HMMSSNR4}{0.217 bits}{HMMSSNR5}{0.353 bits}{HMMSSNR6}{0.486 bits}{HMMSSNR7}{0.666 bits}{HMMSSNR8}{0.769 bits}{HMMSSNR9}{0.856 bits}{HMMSSNR10}{0.919 bits}{HMESNW}{0.000 bits}{HMESSNR1}{0.035 bits}{HMESSNR2}{0.091 bits}{HMESSNR3}{0.217 bits}{HMESSNR4}{0.378 bits}{HMESSNR5}{0.550 bits}{HMESSNR6}{0.699 bits}{HMESSNR7}{0.800 bits}{HMESSNR8}{0.879 bits}{HMESSNR9}{0.924 bits}{HMESSNR10}{0.957 bits}{MMMSNW}{0.000 bits}{MMMSSNR1}{0.006 bits}{MMMSSNR2}{0.009 bits}{MMMSSNR3}{0.017 bits}{MMMSSNR4}{0.013 bits}{MMMSSNR5}{0.026 bits}{MMMSSNR6}{0.021 bits}{MMMSSNR7}{0.070 bits}{MMMSSNR8}{0.167 bits}{MMMSSNR9}{0.199 bits}{MMMSSNR10}{0.350 bits}{MMESNW}{0.000 bits}{MMESSNR1}{0.044 bits}{MMESSNR2}{0.086 bits}{MMESSNR3}{0.174 bits}{MMESSNR4}{0.266 bits}{MMESSNR5}{0.373 bits}{MMESSNR6}{0.444 bits}{MMESSNR7}{0.512 bits}{MMESSNR8}{0.581 bits}{MMESSNR9}{0.619 bits}{MMESSNR10}{0.725 bits}}[\textcolor{red}{???}]}
\DeclareRobustCommand{\VSNRCHIPJS}[1]{\IfEqCase{#1}{{HMMSNW}{0.000 bits}{HMMSSNR1}{0.008 bits}{HMMSSNR2}{0.010 bits}{HMMSSNR3}{0.021 bits}{HMMSSNR4}{0.023 bits}{HMMSSNR5}{0.043 bits}{HMMSSNR6}{0.073 bits}{HMMSSNR7}{0.093 bits}{HMMSSNR8}{0.097 bits}{HMMSSNR9}{0.089 bits}{HMMSSNR10}{0.085 bits}{HMESNW}{0.000 bits}{HMESSNR1}{0.010 bits}{HMESSNR2}{0.016 bits}{HMESSNR3}{0.023 bits}{HMESSNR4}{0.027 bits}{HMESSNR5}{0.028 bits}{HMESSNR6}{0.047 bits}{HMESSNR7}{0.057 bits}{HMESSNR8}{0.125 bits}{HMESSNR9}{0.179 bits}{HMESSNR10}{0.226 bits}{MMMSNW}{0.000 bits}{MMMSSNR1}{0.010 bits}{MMMSSNR2}{0.016 bits}{MMMSSNR3}{0.091 bits}{MMMSSNR4}{0.122 bits}{MMMSSNR5}{0.278 bits}{MMMSSNR6}{0.407 bits}{MMMSSNR7}{0.552 bits}{MMMSSNR8}{0.649 bits}{MMMSSNR9}{0.725 bits}{MMMSSNR10}{0.728 bits}{MMESNW}{0.000 bits}{MMESSNR1}{0.016 bits}{MMESSNR2}{0.027 bits}{MMESSNR3}{0.032 bits}{MMESSNR4}{0.052 bits}{MMESSNR5}{0.055 bits}{MMESSNR6}{0.047 bits}{MMESSNR7}{0.035 bits}{MMESSNR8}{0.012 bits}{MMESSNR9}{0.009 bits}{MMESSNR10}{0.013 bits}}[\textcolor{red}{???}]}

\begin{document}


\title{Inferring the spins of merging black holes in the presence of data-quality issues}


\author{Rhiannon Udall~\orcidlink{0000-0001-6877-3278}}
\email{rhiannon.udall@ubc.ca}
\affiliation{
Department of Physics \& Astronomy, University of British Columbia, Vancouver, BC V6T 1Z1, Canada
}
\affiliation{%
TAPIR, California Institute of Technology, Pasadena, CA 91125, USA
}%
\affiliation{
LIGO Laboratory, California Institute of Technology, Pasadena, California 91125, USA
}
\author{Sophie Bini~\orcidlink{0000-0002-0267-3562}}
\affiliation{%
TAPIR, California Institute of Technology, Pasadena, CA 91125, USA
}%
\affiliation{
LIGO Laboratory, California Institute of Technology, Pasadena, California 91125, USA
}
\author{Katerina Chatziioannou~\orcidlink{0000-0002-5833-413X}}
\email{kchatziioannou@caltech.edu}
\affiliation{%
TAPIR, California Institute of Technology, Pasadena, CA 91125, USA
}%
\affiliation{
LIGO Laboratory, California Institute of Technology, Pasadena, California 91125, USA
}
\author{Derek Davis~\orcidlink{0000-0001-5620-6751}}
\affiliation{Department of Physics, University of Rhode Island, Kingston, RI 02881, USA}
\affiliation{%
TAPIR, California Institute of Technology, Pasadena, CA 91125, USA
}%
\affiliation{
LIGO Laboratory, California Institute of Technology, Pasadena, California 91125, USA
}
\author{Sophie Hourihane~\orcidlink{0000-0002-9152-0719}}
\affiliation{%
TAPIR, California Institute of Technology, Pasadena, CA 91125, USA
}
\affiliation{
LIGO Laboratory, California Institute of Technology, Pasadena, California 91125, USA
}
\author{Yannick Lecoeuche~\orcidlink{0000-0002-9186-7034}}
\affiliation{
Department of Physics \& Astronomy, University of British Columbia, Vancouver, BC V6T 1Z1, Canada
}
\author{Jess McIver~\orcidlink{0000-0003-0316-1355}}
\affiliation{
Department of Physics \& Astronomy, University of British Columbia, Vancouver, BC V6T 1Z1, Canada
}
\author{Simona Miller~\orcidlink{0000-0001-5670-7046}}
\affiliation{%
TAPIR, California Institute of Technology, Pasadena, CA 91125, USA
}
\affiliation{
LIGO Laboratory, California Institute of Technology, Pasadena, California 91125, USA
}

\date{\today}

\begin{abstract}
Gravitational waves from black hole binary mergers carry information about the component spins, but inference is sensitive to analysis assumptions, which may be broken by terrestrial noise transients known as glitches.
Using a variety of simulated glitches and gravitational wave signals, we study the conditions under which glitches can bias spin measurements. 
We confirm the theoretical expectation that inference and subtraction of glitches invariably leaves behind residual power due to statistical uncertainty, no matter the strength (signal-to-noise ratio; SNR) of the original glitch.
Next we show that low-SNR glitches --- including those below the threshold for flagging data-quality issues --- can still significantly bias spin inference.
Such biases occur for a range of glitch morphologies, even in cases where glitches and signals are not precisely aligned in phase.
Furthermore, we find that residuals of glitch subtraction can result in biases as well.
Our results suggest that joint inference of the glitch and gravitational wave parameters, with appropriate models and priors, is required to address these uncertainties inherent in glitch mitigation via subtraction. 
\end{abstract}

\maketitle

\section{Introduction}

In a binary black hole (BBH), the spin magnitudes and orientations provide insights into the processes which generated the BHs and the environment in which the binary exists and evolves~\cite{Fuller:2019sxi,Mandel:2018hfr,Gerosa:2017kvu,Gerosa:2018wbw,Baibhav:2022qxm, Kalogera:1999tq}.
The LIGO-Virgo-KAGRA network of ground-based gravitational-wave (GW) detectors~\cite{LIGOScientific:2014pky,VIRGO:2014yos,KAGRA:2020tym} allows us to characterize these systems with the GWs emitted during their mergers, thus the bulk spin distribution and its outliers are a target of interest for GW observations~\cite{Doctor:2019ruh,Rodriguez:2016vmx,Rodriguez:2019huv,Farr:2017uvj,Zhang:2023fpp,Talbot:2017yur,OShaughnessy:2017eks, Wysocki:2017isg, Callister:2020vyz, Payne:2024ywe,Hannam:2021pit,Tong:2022iws,LIGOScientific:2025jau,Vitale:2015tea, Callister:2022qwb, Miller:2023ncs, Stevenson:2017dlk}. 
A number of BBH events have displayed exceptional spin characteristics, among them GW191109\_010717 (hereafter GW191109) and GW200129\_065458 (hereafter GW200129)~\cite{KAGRA:2021vkt}.
Both originated from heavy BBH systems, with a median total binary mass of 112\,$M_\odot$ and 63\,$M_\odot$, respectively.
Moreover, both signals showed astrophysically interesting spin characteristics, with GW191109 having significant support for spins anti-aligned with the orbital angular momentum and GW200129 having significant spin in the plane of the binary~\cite{KAGRA:2021vkt, Hannam:2021pit, Islam:2023zzj}.
However, both events also experienced anomalous transient noise in the LIGO Livingston detector, which can violate the assumptions under which the above conclusions were drawn~\cite{KAGRA:2021vkt}.
Corrective procedures were applied on the data, but subsequent analyses raised concerns about the impact of terrestrial noise on the spin inference~\cite{Payne:2022spz, Udall:2024ovp}.

Source inference typically assumes that the data are a combination of a transient signal and stationary Gaussian noise~\cite{LIGOScientific:2019hgc,bilby_paper,Cornish:2014kda}. 
These conditions are violated by noise transients, known as ``glitches,'' which can accordingly bias inference~\cite{LIGOScientific:2019hgc, Davis:2022ird,LIGOScientific:2018cki, Cornish:2014kda, Ray:2025rtt}. 
Small perturbations can more easily bias parameters which induce subtle effects in the waveform, such as the system's spins and eccentricity.
When a glitch is coincident with a signal, the typical mitigation approach is to infer the properties of the glitch and subtract one possibility for its true morphology, then proceed as normal~\cite{LIGOScientific:2025yae, Davis:2022ird}; previous studies have validated this process, including most famously in the case of GW170817~\cite{Pankow:2018qpo,Hourihane:2022doe}.
However, due to systematic and statistical uncertainties, it is impossible to perfectly subtract any glitch (or generically, any data component) in noisy data, and residual glitch power could result in biases of subsequent inference.
Furthermore, mitigation is only applied when a glitch has been identified in the data, which is not currently reliably possible for low signal-to-noise-ratio (SNR) glitches, $\rho_g\lessapprox5$~\cite{Robinet:2020lbf, Davis:2022dnd, LIGO:2024kkz}.

In this paper, we perform a series of tests to explore how low-SNR noise artifacts can bias the measurement of spin parameters for massive BBH mergers, focusing on the effective aligned spin $\chieff$~\cite{Racine:2008qv,Ajith:2009bn, Santamaria:2010yb} and the effective precessing spin $\chip$~\cite{Hannam:2013oca, Schmidt:2014iyl}.
To narrow our search space, we consider glitches which overlap with the CBC signal, which \citet{Hourihane:2025vxc} has shown is a prerequisite to biased inference, and which have power concentrated in time-frequency.
We explore glitches with varying SNR, and also with varying time-phase overlap with the CBC signal, which can alter the biases significantly~\cite{Johnson:2024foj}.
We use simulated data and idealized glitch models to investigate the impacts of statistical uncertainties, noting that real data analysis will also be subject to modeling systematics. 

In Sec.~\ref{sec:methodology} we discuss the methods used to model glitches, including a parameterized slow scattering model~\cite{Udall:2022vkv} and a sum of sine-gaussian wavelets~\cite{Cornish:2014kda, Cornish:2020dwh,Chatziioannou:2021ezd, Hourihane:2022doe} model, and CBCs.
We also discuss the configurations~---~characterized by their respective model parameters~---~of the glitches and CBCs which we simulate, which are drawn from the inferred properties of GW191109 and GW200129.

In Sec.~\ref{sec:residualSNR} we investigate the residual SNR associated with traditional glitch subtraction, starting from theoretical expectations~\cite{Cutler:2005qq, Robson:2017ayy}.
We confirm these theoretical expectations using both glitch models and under a variety of analysis settings and choices for drawing point estimates.
Statistical uncertainties result in \textit{residuals with median SNR of 3-7 (depending on the exact model) regardless of the SNR of the original glitch}.
This residual SNR is an inescapable by-product of the glitch subtraction process. 

In Sec.~\ref{sec:varying-glitch-snr} we consider the role a glitch's SNR plays in biasing spin inference, with a particular interest in low-SNR, $\rho_g\lessapprox5$, glitches.
Choosing glitch configurations that are known to have an impact on inference at high SNR, we lower their SNR and simulate data with them overlapping a CBC.
We show that low-SNR glitches can have \textit{significant impacts on the measured CBC spin parameters, including below the threshold of detectability}.

Sec.~\ref{sec:subtraction-biases} builds upon the previous two results by asking whether residuals such as those studied in Sec.~\ref{sec:residualSNR}, which have SNRs comparable to the lower-SNR glitches in Sec.~\ref{sec:varying-glitch-snr}, will also impact spin inference at a comparable level.
We show that in some cases there are substantial biases in the measured spins.
Furthermore, the \textit{presence of biases depends on the exact glitch realization that was subtracted}, e.g., the fair draw from the glitch posterior.
Comparing to analyses which jointly infer the properties of the CBC and glitch, we show that these biases may be addressed when this joint inference is used to marginalize over glitch realizations. 

Finally, in Sec.~\ref{sec:phasetime}, we consider the degree to which these results require precisely aligned overlaps in time and phase between the glitch and the CBC.
We simulate glitches of varying time and phase together with CBCs, and perform standard inference without any mitigation.
We show that \textit{a wide range of glitch parameters can produce significant biases in the CBC spin parameters}, with both the direction and the magnitude of biases varying significantly as a function of the glitch phase.

We conclude in Sec.~\ref{sec:conclusion} by laying out the risks for CBC inference illuminated by this study and discussing alternatives.

\section{Simulation and Inference Methods} \label{sec:methodology}

We use simulated data to explore the statistical uncertainties due to Gaussian noise~---~rather than systematic errors due to model mis-specification~---~for glitch mitigation and subsequent incurred biases on CBC inference. 
Accordingly, in all analyses we use the same models for inference as are used in the creation of the respective data.
This section discusses these models, the simulated configurations, and the algorithms used for inference.

All analyses use four seconds of data with a sampling frequency of $f_\mathrm{samp}=1024\,\mathrm{Hz}$, minimum frequency of $f_\mathrm{min}=20\,\mathrm{Hz}$, and maximum frequency of $f_\mathrm{max}=7/8f_\mathrm{nyq}=7/16f_\mathrm{samp} = 448\,\mathrm{Hz}$ (where $f_\mathrm{nyq}$ is the Nyquist frequency).
For analyses in Sec.~\ref{sec:residualSNR} we use the LIGO design power spectral density (PSD)~\cite{LIGOScientific:2014pky}.
For analyses in Secs.~\ref{sec:varying-glitch-snr}---\ref{sec:subtraction-biases}, we use the PSDs associated with the reference events~\cite{KAGRA:2023pio} as detailed in Table~\ref{tab:simulated-cbcs}.

\subsection{CBC modeling}
\begin{table*}[]
    \begin{tabular}{{|>{\centering\arraybackslash}p{26mm}|>{\centering\arraybackslash}p{13mm}|>{\centering\arraybackslash}p{18mm}|>{\centering\arraybackslash}p{25mm}|>{\centering\arraybackslash}p{26mm}|>{\centering\arraybackslash}p{20mm}|>{\centering\arraybackslash}p{22mm}|>{\centering\arraybackslash}p{16mm}|}} \hline
        \textbf{Configuration name} & \textbf{Abbre-viation} & \textbf{Reference GW event} & \textbf{Detector-frame primary mass $m_1$ ($\mathrm{M}_\odot$)}& \textbf{Detector-frame secondary mass $m_2$ ($\mathrm{M}_\odot$)} & \textbf{Effective aligned spin $\chieff$} & \textbf{Effective precessing spin $\chi_p$} & \textbf{Network Optimal SNR} \\
        \hhline{|=|=|=|=|=|=|=|=|}
        \hmmslong & \hmms & GW191109 & 82 & 65 & 0.05 & 0.31 & 16.1 \\\hline
        \hmeslong & \hmes & GW191109 & 81 & 55 & -0.53 & 0.55 & 16 \\\hline
        \mmmslong & \mmms & GW200129 & 43.1 & 33.6 & 0.2 & 0.27 & 25.2 \\\hline
        \mmeslong & \mmes & GW200129 & 49.5 & 24.4 & -0.02 & 0.97 & 26.5 \\\hline
    \end{tabular}
    \caption{Summary information for the four simulated CBC configurations used in this paper. 
    These are referenced throughout this paper, including in Table~\ref{tab:table-of-runs}, by their configuration name.
    }\label{tab:simulated-cbcs}
\end{table*}

In Table~\ref{tab:simulated-cbcs} we show the four CBC configurations used in this paper.
The \hmmslong{} (\hmms) configuration is drawn from a GW191109 posterior which jointly inferred the parameters of a CBC with the slow scattering model from Eq.~\eqref{eq:slow-scattering} (Run 14 in \citet{Udall:2024ovp}).
Specifically, it is the sample in that posterior which has the highest value of $\chieff=0.05$.
The \hmeslong{} (\hmes) configuration is the maximum likelihood sample from a GW191109 posterior which jointly inferred the CBC with a slow scattering arch (Run 13 in \citet{Udall:2024ovp}), and has strongly negative $\chieff=-0.53$.
The \mmmslong{} (\mmms) configuration is a sample drawn randomly from the a GW200129 posterior produced by \citet{Payne:2022spz} using data in which their ``Case A'' glitch realization is subtracted (Run 10a); it has moderate $\chip=0.27$ and $\chieff=-0.2$.
The \mmeslong{} (\mmes) configuration is the maximum likelihood sample from a GW200129 posterior produced by \citet{Payne:2022spz} using subtracted data (Run 5), and has extremely high $\chip=0.97$.

All CBCs are modeled with the \textsc{NRSur7dq4} waveform approximant~\cite{Varma:2019csw}, which was also used for the analyses from which the above samples are drawn, and which is the most accurate model in its region of validity~\cite{Hannam:2021pit, Varma:2019csw, Payne:2022spz,Islam:2023zzj}.
\textsc{NRSur7dq4} is a surrogate of numerical relativity simulations with mass ratios $m_1/m_2\leq 4$ and component spin magnitudes $\chi_i\leq 0.8$, but may be used in the extrapolation region $m_1/m_2\leq 6$ and $\chi_i \leq 0.99$.
Analyses for \hmms{} and \hmes{} configurations extend into the second extrapolation region, while analyses for \mmms{} and \mmes{} extend into both.
All CBC inference is performed using the \bilby\, inference pipeline using the \textsc{Dynesty} sampler~\cite{bilby_paper, bilby_pipe_paper, Speagle:2019ivv}, with the exception of two joint CBC-glitch inference analyses performed with \bw{}~\cite{Cornish:2014kda, Cornish:2020dwh,Chatziioannou:2021ezd} (Runs~\ref{rid:heavy-moderate-bayeswave-marginalized} and~\ref{rid:moderate-moderate-bayeswave-marginalized} in Table~\ref{tab:table-of-runs}).

\subsection{Data and glitch modeling} \label{sec:glitchesmethods}

We use two models for glitches.
The first is the physically motivated slow scattering model~\cite{Udall:2022vkv, Udall:2024ovp, Tolley:2023umc} which assumes the glitch comes from the scattering of light off of a simple harmonic oscillator.
Under this assumption the phase noise is a sum of sine-of-sines related to each other by fixed frequency intervals
\begin{flalign}\label{eq:slow-scattering}
  \begin{aligned}
    g(t) =& \sum_{k=0}^{N} \biggr{[}A_k \, \times\\&\sin\biggr{(}\frac{f_{h,0}+k\delta f_{h}}{f_{\rm mod}}\sin(2 \pi f_{\rm mod}(t-t_c)) + \phi_k\biggr{)}\biggr{]}\,,
  \end{aligned}
\end{flalign}
where $f_{h,0}$ is the harmonic frequency of the first arch, $\delta f_{h}$ is the frequency spacing between arches, $f_\mathrm{mod}$ is the modulation frequency, $t_c$ is the central time, and the $A_k$ and $\phi_k$'s are amplitudes and phases for the respective arches.
This model is deployed for simulation and inference in Sec.~\ref{sec:scattering-residuals} \runsref{rid:slow-scattering-snr-5}{rid:slow-scattering-snr-50}, with a fixed number of three arches and the \bilby{} inference pipeline.
We use a physical prior configuration which limits the modulating frequency to the microseism band ($0.05~\mathrm{Hz}$---$0.4~\mathrm{Hz}$~\cite{Soni:2023kqq, LIGO:2020zwl}), and place a log-uniform prior on the amplitude.

The second glitch model is the sum-of-wavelets model, which describes glitches as the sum of Morlet-Gabor wavelets~\cite{Cornish:2014kda},
\begin{align}
    g(t) & = Ae^{-\frac{(t-t_0)^2}{\tau^2}}\cos(2\pi f_0(t-t_0) + \phi_0) \nonumber\\
        & = Ae^{-\frac{4\pi^2f_0^2(t-t_0)^2}{Q^2}}\cos(2\pi f_0(t-t_0) + \phi_0)\,,\label{eq:wavelets}
\end{align}
where $f_0$ is the central frequency, $t_0$ is the central time, $A$ is the amplitude, $Q = 2 \pi f_0\tau$ is the quality factor, and $\phi_0$ the phase. 
These wavelets form an overcomplete basis for the space of continuous functions, making them flexible but less informed by the typical properties of true glitches.
The prior distribution for $f_0$ is uniform in [16,512]\,Hz, $Q$ is uniform in [0.1, 40], and $\phi_0$ is uniform in [0, $2\pi$], $t_0$ is uniform in a 4\,s window, the prior on the amplitude of the wavelets is broad, and peaks at SNR=5 per wavelet \cite{Cornish:2020dwh}.
In Sec.~\ref{sec:sum-of-wavelets-residual} the sum-of-wavelets model is sampled with \bw{} in a variety of configurations, including both fixed and variable numbers of wavelets.
We adopt a uniform prior on the number of wavelets.
In Secs.~\ref{sec:varying-glitch-snr}---\ref{sec:subtraction-biases} it is used with \bilby{}, fixing the number of wavelets to one in both simulation and inference.
Inference of glitch parameters may be performed alone, or jointly with the inference of CBC parameters. 
Joint glitch-CBC inference robustly models the statistical uncertainties considered in this paper~\cite{Chatziioannou:2021ezd, Hourihane:2022doe, Malz:2025xdg, Ashton:2022ztk, Plunkett:2022zmx}, as is shown in Section~\ref{sec:subtraction-biases}.

For Secs.~\ref{sec:varying-glitch-snr}---\ref{sec:phasetime} we use glitch configurations which are drawn from wavelet analyses of GW191109 and GW200129 for the high mass and moderate mass cases respectively.
These represent inferences of the possible glitch morphology, which as this work shows does not mean they necessarily reflect the true morphology.
Accordingly, these are plausible representations of true glitches in the Livingston detector, though they do not correspond exactly to any glitch class. 

\subsection{Intuition on glitch-induced bias} \label{sec:predicting-glitch-bias}

In merger-dominated (high-mass) signals, fewer waveform cycles are in the observable frequency band, so individual waveform cycles might drive inference~\cite{Miller:2023ncs,Miller:2025eak}.
Accordingly, the amount of glitch-induced bias should depend on exactly how a glitch overlaps with these cycles~\cite{Udall:2024ovp, Payne:2022spz}.
Under a linear-signal approximation, the bias $\delta \theta^\alpha$ due to a glitch overlapping a CBC signal~\cite{Cutler:2007mi} is quantified as 
\begin{equation}\label{eq:parameter-error-due-to-glitch}
    \delta \theta^\alpha\equiv \hat{\theta}^\alpha - \theta^\alpha_0 \approx (\Gamma^{-1}(\hat{\theta}^{\alpha}))^{\alpha \beta}\langle \mathbf{n} + \mathbf{g}| \partial_\beta \mathbf{h}\rangle\,.
\end{equation}
Here we use $\mathbf{h}$, $\mathbf{n}$ and $\mathbf{g}$ to denote the waveform, Gaussian noise, and glitch respectively, where the bold font indicates these are in data space, i.e. time or frequency.
Furthermore, $\hat{\theta}^\alpha$ are the maximum-likelihood parameters, $\theta^\alpha_0$ are the true parameters, and $\partial_\beta \mathbf{h}$ is the Jacobian of the waveform, with $\beta$ running over the parameters $\theta$.
Finally, $(\Gamma^{-1}(\hat{\theta}^{\alpha}))^{\alpha \beta}$ is the inverse of the Fisher matrix evaluated at the maximum-likelihood parameters,
\begin{equation}
    \Gamma_{\alpha\beta}(\hat\theta) = \langle \partial_\alpha h(\hat\theta)|\partial_\beta  h(\hat\theta)\rangle\,.
\end{equation}
This expression is derived in~\citet{Cutler:2007mi}, except here the cause of bias is a glitch rather than waveform mismodeling. 
Above, $\langle \cdot|\cdot\rangle$ is the noise weighted inner product which has its typical definition~\cite{Cutler:1994ys},
\begin{equation}
    \langle a | b \rangle = 4~\mathrm{Re}\int_{0}^\infty \frac{\tilde a^*(f) \tilde b(f)}{S_n(f)} \text{d} f\,.
\end{equation}
While this formula cannot be used outside of the high-SNR regime, where the linear signal approximation breaks down, it provides intuition for how glitches will impact the inference of CBC parameters.
For discussion of how this bias depends on the sensitivity of the detector, see Appendix~\ref{sec:glitch-bias-sensitivity}.

Notably, Eq.~\eqref{eq:parameter-error-due-to-glitch} shows that the error in an inferred parameter depends on the overlap of the glitch with the \textit{Jacobian} of the waveform model, rather than the waveform itself.
Intuitively, this is because glitches will impact parameter inference not only when they resemble a GW by themselves, but when the sum of the glitch and the CBC signal closely resembles the strain which would be generated by some different CBC configuration.
For a glitch to bias measurements of spins, it only needs to resemble the \textit{difference} between two CBC configurations with different spins, a difference which is often quite subtle for more massive BBH systems. 

\section{What is a glitch residual?} \label{sec:residualSNR}

Just as the presence of Gaussian noise introduces statistical uncertainties in our estimates of astrophysical parameters like BH masses and spins, it also introduces uncertainty in our reconstruction of the glitch.
\textit{The subtraction of any inferred glitch realization will thus result in some non-zero residual glitch SNR.}
Throughout this paper we use the optimal SNR of a (glitch or CBC) template, $\rho =\sqrt{\langle \mathbf{h}|\mathbf{h}\rangle}$, as our measure of the loudness of that template. 

We quantify the residual SNR by simulating a large number of glitches at varying SNRs, inferring them with the methods of Sec.~\ref{sec:methodology}, and then subtracting a glitch realization from the data.
The original ``true" glitches consist of $50$ draws from the glitch prior, scaled to $\mathrm{SNR}\in\{5, 10, 20, 30, 40, 50\}$, and simulated in Gaussian noise.
We then infer posteriors on the parameters of these glitches, and consider the following methods of producing point estimates to subtract: the maximum-likelihood sample, a fair-draw, and the per-frequency median.
The residual SNR
\begin{equation}
    \rho_{\mathrm{res}} = \sqrt{\langle \mathbf{\hat g} - \mathbf{g}_0|  \mathbf{\hat g} - \mathbf{g}_0\rangle}\,,
\end{equation}
is computed for each, where $\mathbf{g}_0$ denotes the true simulated glitch and $\mathbf{\hat g}$ the point-estimate.
By definition $\rho_g\geq 0$, but as we show below, residual SNRs are consistently non-trivial---$\rho_\mathrm{res}\gtrapprox2$ for the simplest models, and greater for more complex models---and typically independent of the SNR of the simulated glitch.

This subtraction problem has also been encountered in the context of the LISA global fit~\cite{Robson:2017ayy, Cutler:2005qq} where the goal is to subtract signals from the data.
When inferring a glitch in Gaussian noise and applying the linear signal approximation, the maximum likelihood configuration may be found with Eq.~\eqref{eq:parameter-error-due-to-glitch}, which takes the form
\begin{equation}\label{eq:parameter-error-due-to-noise}
    \delta \lambda^\alpha \equiv \hat{\lambda}^\alpha - \lambda^\alpha_0\approx (\Gamma(\hat\lambda^\alpha)^{-1})^{\alpha \beta}\langle \mathbf{n} | \partial_\beta \mathbf{g}\rangle\,,
\end{equation}
with the glitch parameters $\lambda$ replacing the CBC parameters $\theta$, $\textbf{g}$ replacing $\textbf{h}$, and only overlap with Gaussian noise being considered. 
From this follows a linear approximation for the residual signal left in the data after subtraction
\begin{equation}
    \delta \mathbf{g} \equiv \mathbf{\hat{g}} - \mathbf{g}_0 \approx \partial_\alpha \mathbf{g}~\delta \lambda^{\alpha}\,.
\end{equation}
Its expectation over noise realizations is~\cite{Cutler:1994ys, Cutler:2005qq, Robson:2017ayy}
\begin{align}
    \mathbb{E}(\langle \delta \mathbf{g}| \delta \mathbf{g}\rangle) &= \mathbb{E}(\langle \partial_\alpha \mathbf{g}~\delta \lambda^{\alpha} | \partial_\beta \mathbf{g}~\delta \lambda^{\beta}\rangle)\nonumber \\
    &=\Gamma_{\alpha\beta}(\mathbf{g}_0)(\Gamma^{-1}(\mathbf{g}_0))^{\alpha\beta}= N_p\,,
\end{align}
where $N_p$ is the number of parameters in the model and $(\Gamma^{-1}(\mathbf{g}_0))^{\alpha\beta}$ is the matrix inverse of $\Gamma_{\alpha\beta}(\mathbf{g}_0)$; the full mathematical derivation is available in the listed references.
We thus expect that 
\begin{equation}\label{eq:res_SNR_maxL}
    \mathbb{E}(\rho_\mathrm{res})\approx\sqrt{N_p}\,.
\end{equation}
In the context of the LISA global fit this relation is applied in the context of signal subtraction, but we show that it holds equally in the case of glitch subtraction in ground based detectors. 

Importantly, this result holds \textit{regardless of the SNR of the true transient} (signal or glitch).
Moreover, as a statistical uncertainty it is present even when there are no systematics due to mis-modeling of the transient. 
Models with greater numbers of parameters, and hence more degrees of freedom, will have greater residual SNR since the added flexibility allows the model to more effectively mimic the Gaussian noise, leading it further astray. 
Perhaps counter-intuitively, the most likely noise realization is the zero-noise relation (the noise distribution is a Gaussian that peaks at zero).
The maximum-likelihood sample is thus the sample which assumes a minimal noise contribution, such that it will usually underestimate the true magnitude of the noise. 
Similar considerations apply to the other point-estimates, as they can never exactly match the true glitch.

\subsection{Slow scattering model} \label{sec:scattering-residuals}

\begin{figure}
    \centering
    \includegraphics[width=0.49\textwidth]{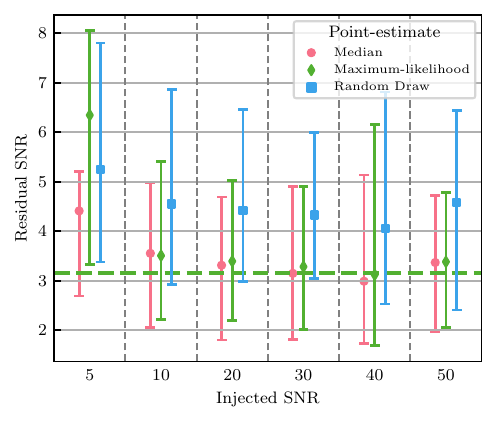}
    \caption{Residual SNR after subtracting a point-estimate for the slow scattering glitch model as a function of the injected glitch SNR, over the range $\rho_g \in \{5, 10, 20, 30, 40, 50\}$. 
    The three point-estimates---the maximum-likelihood point, a random draw from the posterior, and the frequency point-wise median of the posterior---are shown in green, blue, and pink respectively, and each grouping corresponds to the same set of injections and recoveries.
    The markers represent median values, while the whiskers represent 90\% intervals over 50 glitches simulated from the prior.
    The green dashed line is placed at $\sqrt{10}$, the theoretical prediction for the expectation value of the residual SNR after subtracting the maximum-likelihood estimate; see Eq.~\eqref{eq:res_SNR_maxL}.
    }
    \label{fig:slow-scattering-residual-snr}
\end{figure}

We begin with the parameterized slow scattering model, using a configuration with three scattering arches.
Simulated glitches are drawn from the prior distribution and then the arch amplitudes are scaled uniformly to adjust the SNR.
In Fig.~\ref{fig:slow-scattering-residual-snr}, we plot the residual SNR for the three choices of point-estimate to subtract and as a function of the simulated (injected) SNR.
The model has $N_p = 10$, and so Eq.~\eqref{eq:res_SNR_maxL} predicts a residual SNR of $\mathbb{E}(\rho_\mathrm{res}) \approx 3.16$, illustrated with a horizontal dashed line. 

The residual SNR is largely independent of the injected SNR for $\rho_{\mathrm{inj}} \gtrapprox 5$, even for high glitch SNRs.
The maximum-likelihood result (green) generally aligns with the theoretical prediction, while the residual SNR for median glitch realizations is consistent, albeit slightly lower.
The SNR of residuals from posterior random draws are substantially higher and have a larger range of values, likely due to the stochastic nature of sampling. 
The case of $\rho_{\mathrm{inj}}=5$ is unique, as the linear signal approximation, and thus Eq.~\eqref{eq:res_SNR_maxL}, no longer holds.
Now residual SNRs are frequently \textit{higher than the SNR of the original glitch}, which we interpret as a failure to identify the glitch in the first place when SNRs are so low.
This is concerning, since it implies that \textit{for sufficiently low-SNR glitches, attempting subtraction may be doing more harm than good}.
Furthermore, that assumes the glitches are even identified at all, since tools such as \textsc{Omicron} lose sensitivity at $\rho_g \lessapprox 5$~\cite{Robinet:2020lbf}.
As we shall see in Sec.~\ref{sec:varying-glitch-snr}, such low-SNR glitches can still impact parameter inference results. 

\subsection{Sum-of-wavelets model} \label{sec:sum-of-wavelets-residual}

\begin{figure*}
 \centering
 
\includegraphics[width=0.9\linewidth]{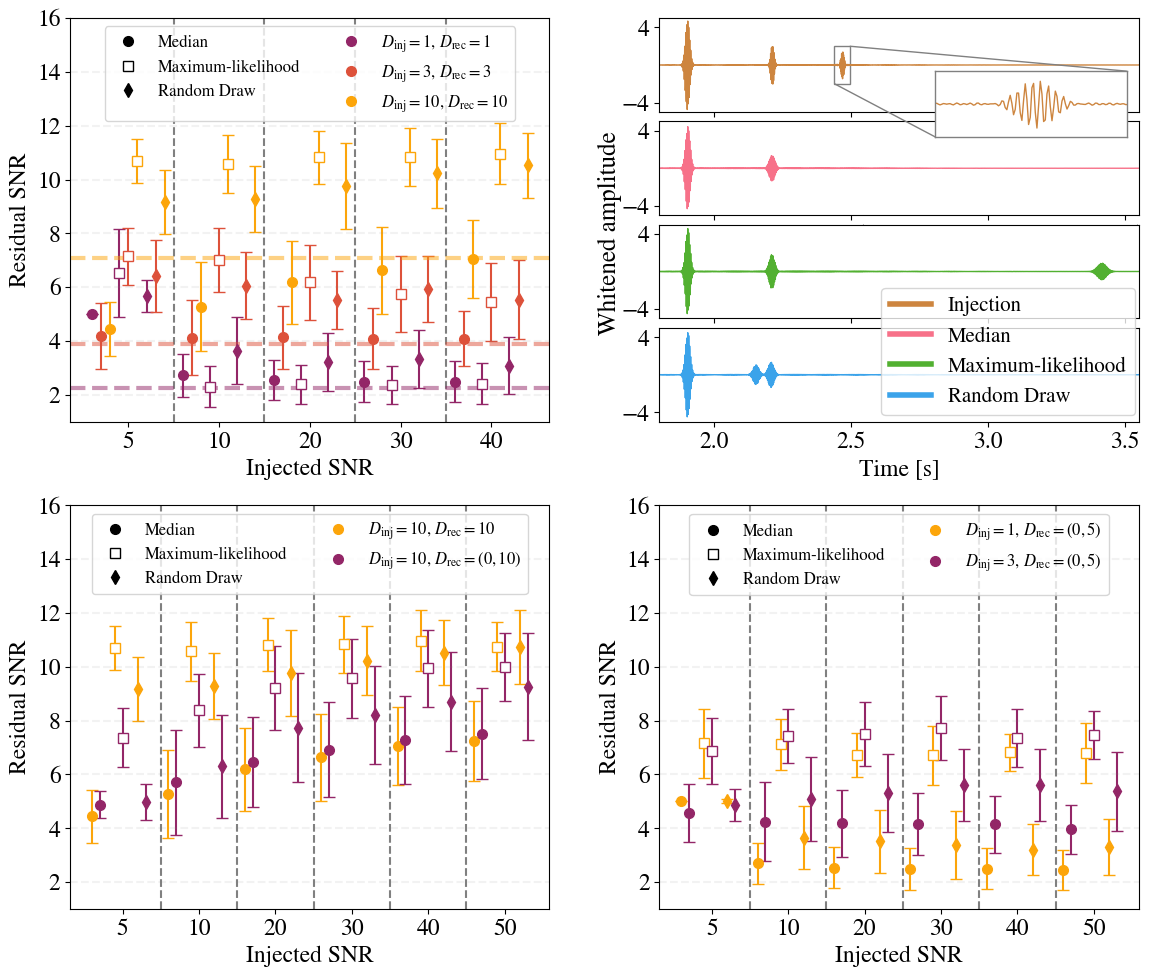}
\caption{Similar to Fig.~\ref{fig:slow-scattering-residual-snr} but when the glitch is as a sum-of-wavelets modeled with \bw\,(Eq.~\eqref{eq:wavelets}). \textit{Top Left}: The number of injected and recovered wavelets is the same and fixed to  $D_\mathrm{rec}=D_\mathrm{inj}=1,3,10$. This panel explores the impact of increasing model complexity.
\textit{Top Right:} An example from the $D_\mathrm{rec}=D_\mathrm{inj}=3$ at $\rho_{\mathrm{inj}}=20$ case. We plot the whitened waveforms generated from the injected, median, maximum-likelihood and random posterior draw parameters. 
\textit{Bottom Left}: The number of injected wavelets is $D_\mathrm{inj}=10$ and we recover it with either fixing the number of wavelets to $D_\mathrm{rec}=10$ or varying $D_\mathrm{rec}=(0,10)$. This panel explores the impact of variable-dimensionality inference. 
\textit{Bottom Right:} We present a more realistic scenario where 1 and 3 injected wavelets are modeled with a flexible model with variable dimensionality $D_\mathrm{rec}=(0,5)$.
The number of maximum wavelets possibly included in the model is higher than the number of injected components, reflecting our lack of knowledge when analyzing real data.
}
\label{fig:BW_residual}
\end{figure*}

Next, we investigate glitch subtraction using \bw{}'s model, which is a sum of a variable number of wavelets (Eq.~\eqref{eq:wavelets}).
This model's flexibility allows it to target a wide range of glitch morphologies overlapping with real signals~\cite{KAGRA:2021vkt}. 
We investigate both a fixed and a varying number of wavelets, thus exploring models with increasing complexity and realism. 
We again simulate glitches from the model prior and rescale them to a desired SNR. 
Below, the number of wavelets in simulated glitches is denoted $D_\mathrm{inj}$, while the number of wavelets allowed during inference is denoted $D_\mathrm{rec}=\cdot$ and $D_\mathrm{rec}=(\cdot, \cdot)$ for fixed- and varying-dimensionality inference respectively.
In the latter case, parenthesis denote the minimum and maximum number allowed. Results are shown in Fig.~\ref{fig:BW_residual} in a similar format as Fig.~\ref{fig:slow-scattering-residual-snr}.

As in the slow scattering case (Fig.~\ref{fig:slow-scattering-residual-snr}), for $\rho_{\mathrm{inj}}=5$ the residual SNRs can be higher than the SNR of the injected glitch.
This again indicates that such low-SNR glitches might not be identified in the first place, despite the fact that they are inconsistent with Gaussian noise.
The following discussion focuses on the simulations with $\rho_{\mathrm{inj}}\geq 10$.

We begin with the case of injection and recovery with the same, fixed number of wavelets.
The top left panel of Fig.~\ref{fig:BW_residual} shows results for $D_\mathrm{inj}=D_\mathrm{rec}=1, 3, 10$, corresponding to increasing complexity. 
Again, the residual SNR does not depend on the injected SNR.
In the simplest case, $D_\mathrm{inj}=D_\mathrm{rec}=1$ (purple), we obtain similar results as the low scattering model of Fig.~\ref{fig:slow-scattering-residual-snr}: the maximum-likelihood and median residual SNRs are consistent with the expectation from Eq.~\eqref{eq:res_SNR_maxL} (${\sim} 2.5\approx\sqrt{5}\approx2.2$), while the random draw residual SNR is ${\sim} 3.2$, so higher.

The residual SNR is expected to increase with the complexity of the model, $\sqrt{N_p}=\sqrt{5 D_\mathrm{inj}}$.
For more complex glitches, $D_\mathrm{inj}=D_\mathrm{rec}=3$ (pink) and especially $D_\mathrm{inj}=D_\mathrm{rec}=10$ (orange), the residual SNR from the median estimate generally aligns with this expectation, however both the maximum-likelihood and random draw residuals are even higher.
This is caused by the limitations of modeling complex glitches: since we rescale the wavelet amplitudes to achieve a desired SNR, even at $\rho_{\mathrm{inj}}=50$ some wavelets will have a low SNR and thus be below the detectability threshold.
As $D_\mathrm{rec}=D_\mathrm{inj}$, the model is nonetheless forced to use wavelets. 
But if the SNR of an injected wavelet is low, the model wavelets may not be correctly recovering it. 
As shown in the top right panel, such ``spurious" low-SNR wavelets are averaged out in the median, but will increase the residual SNR in the maximum likelihood and the random draw.
Even though glitches in real data do not contain spurious low-SNR wavelets, they still contain complicated low-SNR features that are subject to the same consideration.
In other words, the Fisher approximation assumes that the posterior for \textit{all} parameters is Gaussian, which might not be true even at seemingly very high total SNRs.
These considerations suggest that 1. \textit{the median inferred glitch is the most reliable point estimate} and 2. \textit{artificially increasing model complexity in an attempt to capture low-SNR features is counterproductive}.

Balancing the above concerns about needlessly complex models while being able to model complex glitches,
\bw{} uses a trans-dimensional glitch model. 
Depending on the glitch complexity, the model can transition between solutions with different number of wavelets.  
We explore the effect of the trans-dimensionality by recovering the signals of the top left panel with $D_\mathrm{inj}=10$ with fixed $D_\mathrm{rec}=10$ and variable $D_\mathrm{rec}=(0,10)$.
The bottom left panel of Fig.~\ref{fig:BW_residual} shows that when subtracting the median, the residual SNR is similar under both a fixed or a variable number of wavelets, as low-SNR wavelets are averaged out as explained above. 
Instead, when subtracting the maximum-likelihood or the random draw, the residual SNR decreases under variable $D_\mathrm{rec}$.
This is because the trans-dimensional model is not forced to add wavelets in a (potentially failed) attempt to identify low-SNR features.
However, the lowest residual SNR is achieved with the median subtraction which remains the preferred option: ${\sim 5.3}$ at $\rho_{\mathrm{inj}}=10$, and ${\sim}7.2$ at $\rho_{\mathrm{inj}}=40$.
Interestingly, while the maximum-likelihood residual SNR does not depend on the injected SNR when $D_\mathrm{rec}$ is fixed, it has an upward trend when $D_\mathrm{rec}$ is variable as the model adds wavelets at higher SNRs.

In real data, of course, we have no advance knowledge of the number of wavelets needed.
The standard procedure is to employ a varying-dimensionality model with a sufficiently large maximum number of wavelets such that the prior does not restrict the posterior \cite{Cornish:2014kda}. 
In the bottom right panel, we inject a simple glitch with $D_\mathrm{inj}=1$ (e.g., a blip glitch \cite{Cabero:2019orq}) and a more complex glitch $D_\mathrm{inj}=3$, and model them with $D_\mathrm{rec}=(0,5)$. 
As expected again, the residual SNR does not depend on the injected SNR. 
The lowest residual SNR is again from subtracting the median, ${\sim}2.5$ for $D_\mathrm{inj}=1$ and ${\sim}4.2$ for $D_\mathrm{inj}=3$, while the highest residual SNR is from subtracting the maximum-likelihood estimate, ${\sim}6.8$ for $D_\mathrm{inj}=1$ and ${\sim}7.5$ for $D_\mathrm{inj}=3$.
Even though the recovery model is the same, the more complex injection has a higher SNR residual: as discussed above, low-SNR features in the glitches are difficult to model and subtract.

\section{The role of glitch SNR in spin inference biases} \label{sec:varying-glitch-snr}

The impact of high-SNR, flagged glitches on CBC inference is well-documented~\cite{Pankow:2018qpo,Chatziioannou:2021ezd,Hourihane:2022doe, Payne:2022spz, Udall:2024ovp, Davis:2022ird, Ghonge:2023ksb}.
Here, we consider subthreshold glitches that might not have been flagged, $\rho_g\lessapprox5$. 
Such glitches with low SNRs are likely present in the data, but we do not have reliable means of detecting them. 
If they impact inference, biases may be possible \textit{without us even knowing that a glitch is present in the data}.

For each CBC signal of Table~\ref{tab:simulated-cbcs}, we identify a glitch configuration that incurs a bias, see subsequent Sec.~\ref{sec:phasetime}. 
We then simulate data with the CBC and the glitch in zero-noise at various SNRs and infer the CBC properties in the standard way, ignoring the glitch. 
We quantify parameter biases in two ways. First, via the posterior credible level of the true value for $\chieff$ and $\chip$, and second, with the Jensen-Shannon (JS) divergence (in bits) between these posteriors and their respective reference cases in which no glitch is added.
We quote credibility by the percentage of samples which have probability density less than that of the true value, such that 100\% corresponds to the case where the true value is the most probable value, and 0\% corresponds to the case where the true value is excluded by the posterior.  
For the JS divergence we compare to three thresholds: $\mathrm{JS}=0.007, 0.044$, and $0.161$ bits.\footnote{
\citet{LIGOScientific:2020ibl} adopted a threshold of $\mathrm{JS}\geq0.007~\mathrm{bits}$ to designate when two posteriors were in disagreement, which we consider to be exceptionally conservative.
This number is the JS divergence of two Gaussian distributions of the same standard deviation and offset in their mean by $\sigma/5$.
This setup is somewhat deceptive because if the mean is shifted by $\delta \mu = a\sigma$, then the JS divergence is quadratic in $a$ for $a\lessapprox1$.
The other reference values we adopt, $0.044$ and $0.161$, are obtained for $a=0.5$ and $a=1$, illustrating the quadratic progression.
Qualitatively these three thresholds span the distance from noticeable differences to severe biases.
}

\begin{figure*}
    \centering
    \includegraphics[width=\textwidth]{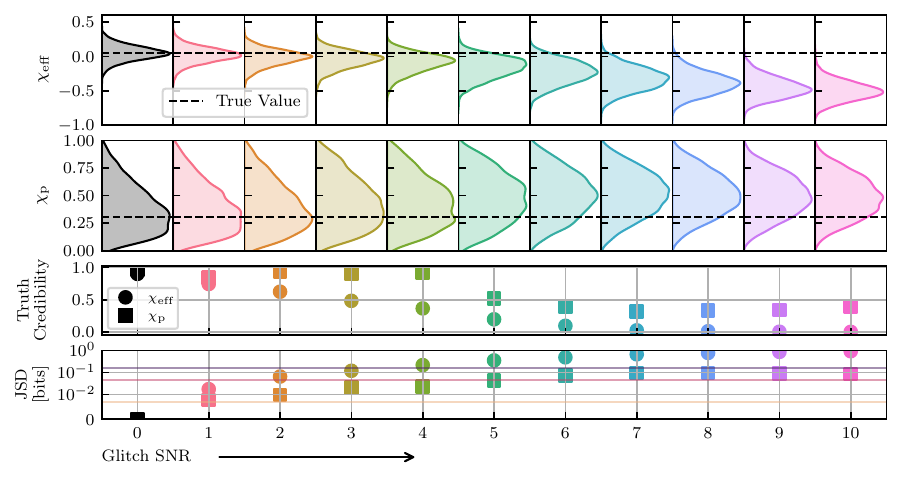}
    \caption{Analyses of the \hmms{} CBC configuration \runsref{rid:heavy-moderate-snr-start}{rid:heavy-moderate-snr-end} simulated along with a glitch of varying SNR from 0 (i.e., the reference CBC-only case) to 10.
    The top panels show posteriors for $\chieff$ (top) and $\chip$ (second from top) as a function of the glitch SNR.
    True values are marked with a black dashed line.
    The bottom panels show the credibility of the true value and the JS divergence of the given posterior against the CBC-only posterior, with circles corresponding to $\chieff$ and to $\chip$.
    The bottom panel has three lines at JS divergences of $0.007, 0.044$, and $0.161$, corresponding to the three JS divergence thresholds we compare against.
    }\label{fig:heavy-moderate-varying-snr}
\end{figure*}

\begin{figure*}
    \centering
    \includegraphics[width=\textwidth]{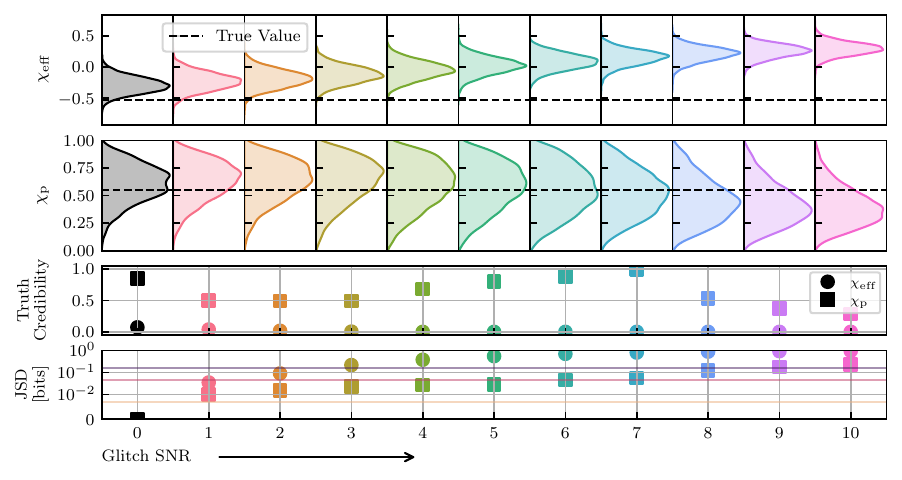}
    \caption{Same as Fig.~\ref{fig:heavy-moderate-varying-snr} for the \hmes{} CBC configuration \runsref{rid:heavy-extreme-snr-start}{rid:heavy-extreme-snr-end} simulated along with a glitch of varying SNR.
    }\label{fig:heavy-extreme-varying-snr}
\end{figure*}

\begin{figure*}
    \centering
    \includegraphics[width=\textwidth]{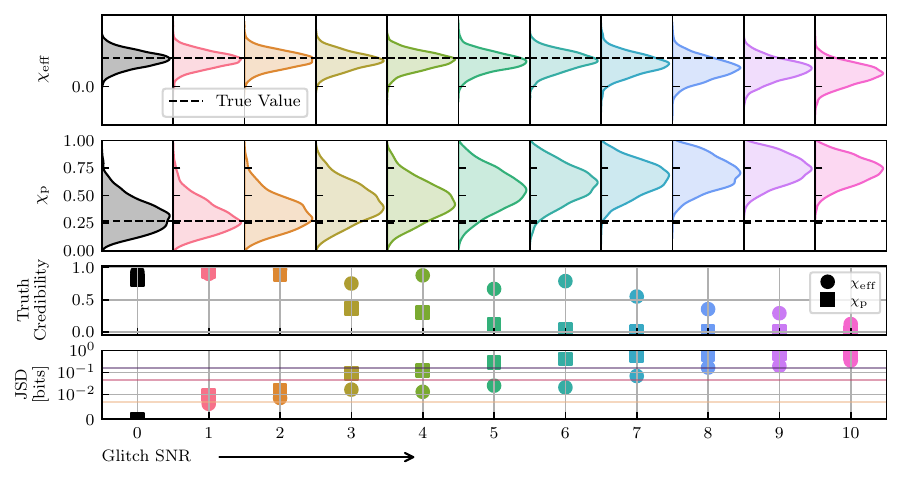}
    \caption{Same as Fig.~\ref{fig:heavy-moderate-varying-snr} for the \mmms{} CBC configuration \runsref{rid:moderate-moderate-snr-start}{rid:moderate-moderate-snr-end} simulated along with a glitch of varying SNR.
    }\label{fig:moderate-moderate-varying-snr}
\end{figure*}

\begin{figure*}
    \centering
    \includegraphics[width=\textwidth]{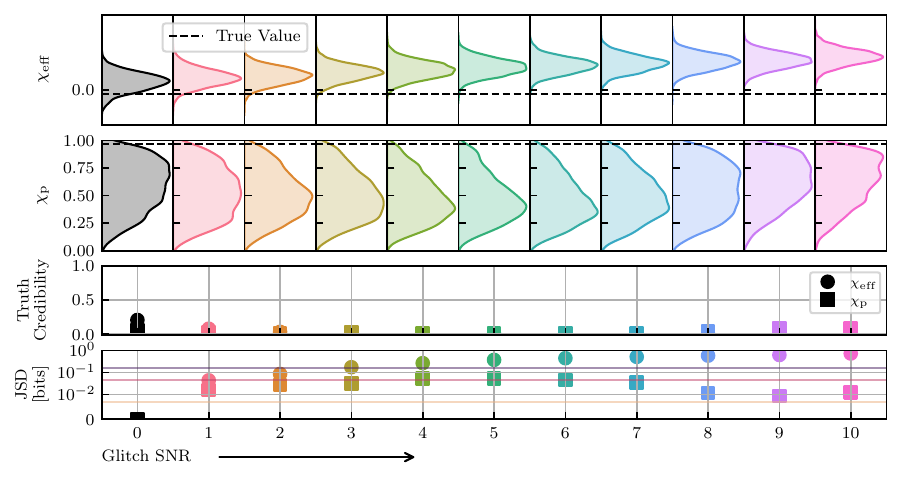}
    \caption{Same as Fig.~\ref{fig:heavy-moderate-varying-snr} for the \mmes{} CBC configuration \runsref{rid:moderate-extreme-snr-start}{rid:moderate-extreme-snr-end} simulated along with a glitch of varying SNR.
    }\label{fig:moderate-extreme-varying-snr}
\end{figure*}

Figure~\ref{fig:heavy-moderate-varying-snr} shows results for the \hmms{} CBC configuration simulated along with a glitch of varying SNR \runsref{rid:heavy-moderate-snr-start}{rid:heavy-moderate-snr-end}. 
Posteriors for the reference (CBC-only) analysis are peaked at the true value \runref{rid:zero-heavy-moderate}; inclusion of the glitch causes a significant shift in the posterior towards negative $\chieff$ and smaller shifts in $\chip$.
The bias in $\chieff$ at $\rho_g=10$ is severe---the true value has \VSNRCHIEFF{HMMSSNR10} credibility and the JS divergence with the CBC-only posterior is \VSNRCHIEFFJS{HMMSSNR10}---as one would expect from such a loud glitch.
More worringly, the bias in $\chieff$ at $\rho_g=5$ is already significant with the true value having \VSNRCHIEFF{HMMSSNR5} credibility and the JS divergence with the CBC-only posterior being \VSNRCHIEFFJS{HMMSSNR5}.
The latter exceeds our most stringent threshold for concern, even at a glitch SNR which likely would make it undetectable.
According to the most conservative criterion of $\mathrm{JS}\geq0.007$, even the $\rho_g=1$ posterior (\VSNRCHIEFFJS{HMMSSNR1}) is in disagreement with the reference distribution; this will be true of at least one of $\chieff$ or $\chip$ in each of the cases we consider below.

Figure~\ref{fig:heavy-extreme-varying-snr} shows results for the \hmes{} CBC configuration \runsref{rid:heavy-extreme-snr-start}{rid:heavy-extreme-snr-end}. 
Here the reference \runref{rid:zero-heavy-extreme} posterior in $\chieff$ peaks nearer 0 than the true value, which has credibility \VSNRCHIEFF{HMESNW}, and the glitch causes shifts towards more positive values of $\chieff$.
As with the \hmms{} case, the $\chieff$ bias for $\rho_g=10$ is severe, with the true value having \VSNRCHIEFF{HMESSNR10} credibility and JS divergence \VSNRCHIEFFJS{HMESSNR10}.
Also similar to the \hmms{} case, severe biases start at lower SNRs, with the $\rho_g=3$ case having the true value at \VSNRCHIEFF{HMESSNR3} credibility and JS divergence \VSNRCHIEFFJS{HMESSNR3}, surpassing the most stringent JS threshold for a bias.
In this case the glitch is making the $\chieff$ posterior more moderate, i.e., less negative, than the true value.
Qualitatively similar biases are also present in $\chip$, where the posterior is broad, but shifts around due to the glitch.

Next, Fig.~\ref{fig:moderate-moderate-varying-snr} shows results for the \mmms{} configuration \runsref{rid:moderate-moderate-snr-start}{rid:moderate-moderate-snr-end}, which suffers from severe biases in $\chip$.  
The reference \runref{rid:zero-moderate-moderate} posterior has the true $\chip$ value comfortably at \VSNRCHIP{MMMSNW} credibility.
For a glitch with $\rho_g=5$, however, the true value's credibility has fallen to \VSNRCHIP{MMMSSNR5} and the JS divergence with the CBC-only case is \VSNRCHIPJS{MMMSSNR5}.
Posteriors for $\chieff$ show less bias for these middling SNRs, with the $\rho_g=5$ case having the true $\chieff$ value at \VSNRCHIEFF{MMMSSNR5} credibility and a JS divergence of \VSNRCHIEFFJS{MMMSSNR5} against the CBC-only case.
However, for sufficiently high SNRs they too show significant biases, with the true value's credibility being reduced to \VSNRCHIEFF{MMMSSNR10} when the glitch has $\rho_g=10$.

Finally, Fig.~\ref{fig:moderate-extreme-varying-snr} shows results for the \mmes{} configuration \runsref{rid:moderate-extreme-snr-start}{rid:moderate-extreme-snr-end}.
As with the \hmes{} case, the extreme spins in regions of low prior support---here $\chi_p=0.97$---result in posteriors that do not peak at the true value; here only at \VSNRCHIP{MMESNW} credibility.
Among the cases we have considered this one is unique: the $\chip$ posterior is \textit{worse} for the $\rho_g=5$ case, where the true value is rejected with \VSNRCHIP{MMESSNR5} with a JS divergence against the CBC-only case of \VSNRCHIPJS{MMESSNR5}, than the $\rho_g=10$ case, where the true value has \VSNRCHIP{MMESSNR10} credibility and a JS divergence of \VSNRCHIPJS{MMESSNR10}.
The posteriors in $\chieff$ shift in a manner more typical of the other cases, with the glitch with $\rho_g=5$ having the true value at \VSNRCHIEFF{HMESSNR5} credibility and JS divergence of \VSNRCHIEFFJS{HMESSNR5} against the posterior, and the glitch with $\rho_g=10$ having the true value at \VSNRCHIEFF{HMESSNR10} credibility and JS divergence of \VSNRCHIEFFJS{HMESSNR10} against the CBC-only case.


The most striking conclusion is consistent across all cases: \textit{glitches with $\rho_g\leq5$---which will very likely go undetected---can have substantial impacts upon the inference of spin parameters}.
As we show later in Sec.~\ref{sec:subtraction-biases}, this pattern will also hold for glitch residuals of similar SNRs.
The posterior shifts do not happen linearly with the glitch SNR, and are not even necessarily monotonic
While Gaussian noise realizations may also produce significant shifts in the inferences of spins, the shifts here occur systematically and, unlike Gaussian noise, they are not accounted for in the assumptions made by current analyses.  

\section{Spin inference biases due to point-estimate subtraction} \label{sec:subtraction-biases}

\begin{figure*}
    \centering
    \includegraphics[width=\textwidth]{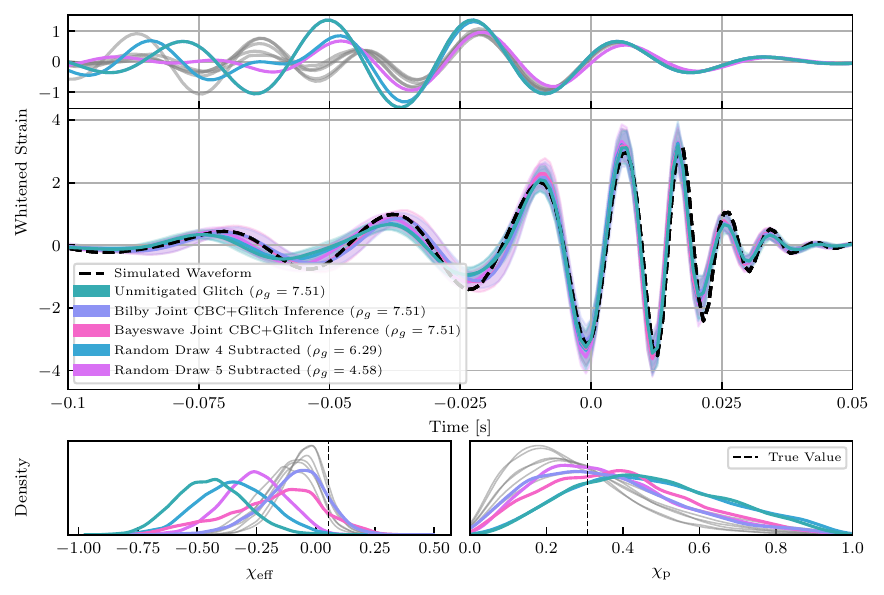}
    \caption{
    A \hmms{} simulated with a glitch \runsref{rid:gaussian-heavy-moderate}{rid:heavy-moderate-marginalized}, with point estimates subtracted using the corresponding glitch-only inference \runref{rid:heavy-moderate-wavelet-inference}.
    Eleven cases are present: Gaussian noise only, the glitch added without any mitigation applied, two analyses performing joint glitch-CBC inference with \bilby{} and \bw{} respectively, and seven cases in which realizations (median, maximum likelihood, and five samples drawn randomly from the posterior) are subtracted. 
    The top panel shows the glitch---including both the original glitch and various residuals after subtraction---which were added on top of the simulated CBC and Gaussian noise, with notable cases highlighted and other realizations in gray.
    The middle panel shows posterior CBC reconstructions (median and 90\% credible intervals) for the notable cases, and the true simulated CBC as a gray dashed line. 
    The bottom panels show posteriors in $\chieff$ (left) and $\chip$ (right), once again highlighting notable cases while leaving all other cases in gray, and marking the true value with a black dashed line.
    }
    \label{fig:heavy-moderate-subtractions}
\end{figure*}

\begin{figure*}
    \centering
    \includegraphics[width=\textwidth]{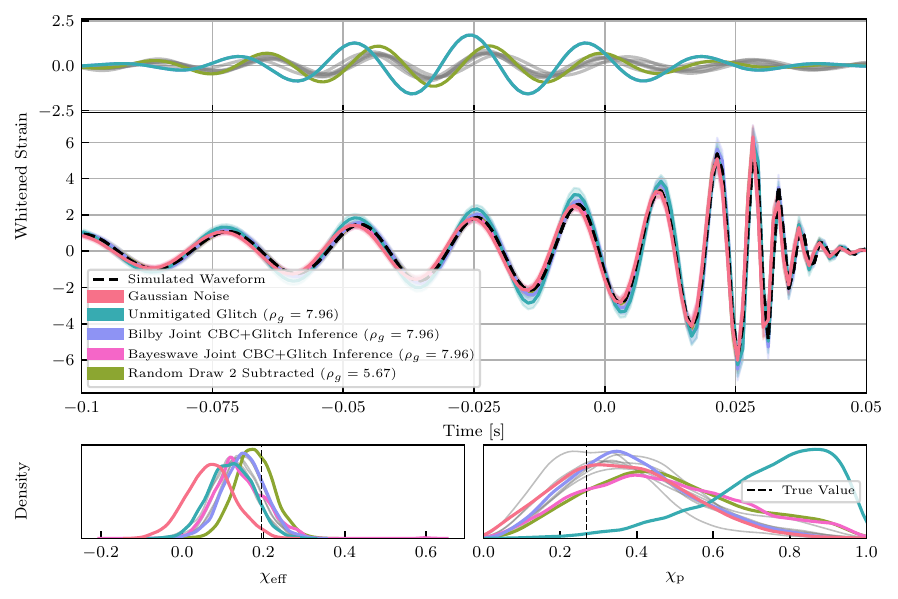}
    \caption{Same as Fig.~\ref{fig:heavy-moderate-subtractions} for the \mmms{} CBC with a glitch \runsref{rid:gaussian-moderate-moderate}{rid:moderate-moderate-marginalized}, with point estimates subtracted using the corresponding glitch-only inference \runref{rid:moderate-moderate-wavelet-inference}.
    The same set of cases appear as in  Fig.~\ref{fig:heavy-moderate-subtractions}, but different cases are highlighted.
    }
    \label{fig:moderate-moderate-subtractions}
\end{figure*}

Since Sec.~\ref{sec:residualSNR} shows that subtraction with point-estimates can leave residuals with non-trivial SNRs of $\rho_g \gtrapprox 3$ and Sec.~\ref{sec:varying-glitch-snr} shows that glitches with SNRs in this range may produce significant biases in spin inference, we now turn to the question of whether these \textit{glitch residuals} also result in biases. 
The morphology of the residual will generally not be the same as the original glitch, and so the effects of residuals on inference may differ from that of unsubtracted glitches with comparable SNRs. 
They will also depend on the choice of point-estimate. 

In order to test the above, we perform a two-step analysis roughly analogous to the inference-subtraction procedure used for most GW inference studies.
First, we simulate glitches as a single wavelet in Gaussian noise (using the same noise realization for each test to isolate the effects of the glitch), using parameters that result in significant bias when overlapping with CBC signals (see Appendix~\ref{sec:gaussian-phase-scan} for details), and infer the glitch parameters using~\textsc{bilby}.
We then obtain glitch realizations using the variety of methods described in Sec.~\ref{sec:residualSNR}: median, maximum-likelihood, and five random posterior draws.
We subtract them from the data, add a CBC, and analyze the data in the standard way with a CBC model only.
The data are now a combination of the CBC, the glitch residual, and the underlying Gaussian noise.
This procedure is a best-case scenario when compared to realistic glitch subtraction procedures, because in that case it is necessary to infer the glitch using data which already includes the CBC, potentially introducing correlations.
For comparison, we also analyze the full-glitch and CBC data with joint glitch-CBC inference, once using \bilby{} and modeling the glitch with only one wavelet and then again using \bw{} with a variable number of wavelets.

Figure~\ref{fig:heavy-moderate-subtractions} shows the results of this procedure for the \hmms{} CBC system~\runsref{rid:moderate-moderate-maxL-subtraction}{rid:moderate-moderate-bayeswave-marginalized}, and a glitch with original SNR ${\sim}7.5$ \runref{rid:heavy-moderate-wavelet-inference}. 
We highlight five cases: no glitch subtraction, joint CBC-glitch inference using \bilby{} and \bw{}, and two random-draw residuals that generated significant biases.
Naturally, the biases are worst for the case where no glitch mitigation is employed, which assigns the true value of $\chieff$ at \RICHIEFF{HMMSUS} credibility.
Less severe, but still significant, are the biases in $\chieff$ for two (out of five) of the randomly drawn point-estimates; these both assign the true value at \RICHIEFF{HMMSFD4} credibility.
The \bilby{} joint CBC-glitch analysis with fixed glitch dimensionality effectively mitigates biases and returns results consistent with the true value in $\chieff$, assigning it \RICHIEFF{HMMSJI} credibility.
The \bw{} joint CBC-glitch analysis which allows the dimensionality of the glitch model to vary also mitigates biases, finding the true value at \RICHIEFF{HMMSBW}, but yields a posterior with a long tail towards negative $\chieff$ values.
This is an example of a correlation between the CBC and the glitch model parameters: the long tail corresponds to no glitch wavelets in the posterior, meaning that for this configuration, a CBC with more extreme $\chieff$ has a similar morphology to the sum of the glitch and a moderate-$\chieff$ CBC.
The no-subtraction case (cyan) and the fifth randomly sampled residual (blue) also see a shift in $\chip$, though the posterior remains prior-dominated such that the credibility of the true value of $\chip$ only drops from \RICHIP{HMMSJI} to \RICHIP{HMMSUS} between the jointly inferred and no-subtraction cases respectively. 

Figure~\ref{fig:moderate-moderate-subtractions} shows corresponding results for the \mmms{} CBC configuration, and a glitch with original SNR ${\sim}8.0$ \runref{rid:moderate-moderate-wavelet-inference}.
Once again we highlight five cases of note: no glitch subtraction, joint CBC-glitch inference using \bilby{} and \bw{}, no glitch (i.e., only the CBC and Gaussian noise), and one of the random draws.
By far the worst biases in $\chip$ are found for the case with no mitigation applied, which has the true value at \RICHIP{MMMSUS} credibility.
The other cases have relatively limited biases in $\chip$, finding the true value at \RICHIP{MMMSJI}, \RICHIP{MMMSBW}, \RICHIP{MMMSGN}, and \RICHIP{MMMSFD2} credibility for the joint CBC-glitch inference with \textsc{bilby}, the joint CBC-glitch inference with \bw{}, Gaussian noise only, and the random draw respectively. 
The $\chieff$ inference is more complicated, with the worst measurement being the case with no glitch at all, which finds the true value at only \RICHIEFF{MMMSGN} credibility.
Meanwhile the other cases in which a glitch is present and possibly mitigated find the true value at \RICHIEFF{MMMSUS}, \RICHIEFF{MMMSJI}, \RICHIEFF{MMMSBW}, and \RICHIEFF{MMMSFD2} credibility for the cases with no glitch mitigation applied, joint CBC-glitch inference with \textsc{bilby}, joint CBC-glitch inference with \bw{}, and subtraction of the randomly sampled realization respectively. 
This situation is harder to characterize: evidently due to Gaussian noise fluctuations the true value was disfavored, but the presence of any glitch at all causes the posteriors to push more towards the true value, while still remaining consistently lower than the truth. 
The lack of significant residual biases such as appeared for the \hmms{} case is likely attributable to luck-of-the-draw. 

\section{The effect of glitch time and phase} \label{sec:phasetime}

\begin{figure*}
    \centering
    \includegraphics[width=\textwidth]{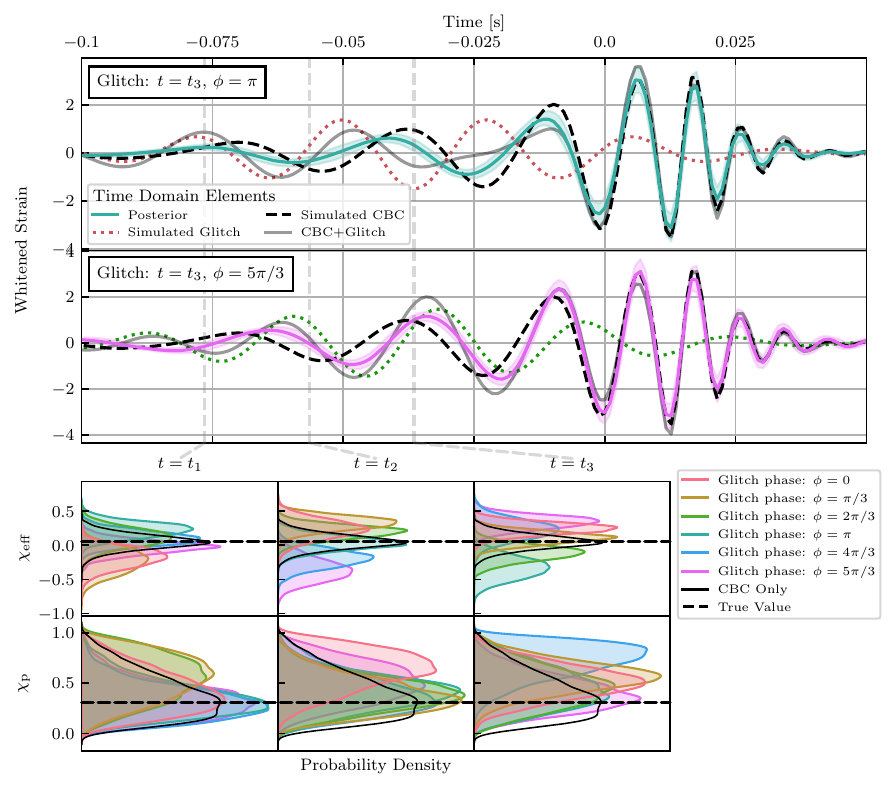}
    \caption{Analyses of the \hmms{} configuration simulated along with glitches of varying time and phase \runsref{rid:zero-heavy-moderate-phase-start-t1}{rid:zero-heavy-moderate-phase-end-t3}.
    The two top panels show data and posterior reconstructions for two highlighted examples: the analysis with a glitch at $t=t_3, \phi=\pi$ (top panel) and $t=t_3, \phi=5\pi/3$ (second panel).
    For these panels, the dashed blue line shows the original simulated CBC signal, the dotted line (red in the top, green in the bottom) shows the simulated glitch, the solid black line shows the sum of the two, and the colored solid line with its corresponding shaded region (green in the top, pink in the bottom) shows the posterior reconstruction median and 90\% credible interval. 
    Light gray dashed vertical lines illustrate the three central times labeled $t_1, t_2$, and $t_3$, and connect them to respective posterior panels containing results for that time. 
    The bottom panels show posteriors in $\chieff$ and $\chip$ for all of these times (first three left to right), as well as for a reference analysis with no glitch simulated in the rightmost panel. 
    They also show the true value of each parameter marked with a black dashed line.
    }\label{fig:heavy-moderate-phase-scans}
\end{figure*}

\begin{figure*}
    \centering
    \includegraphics[width=\textwidth]{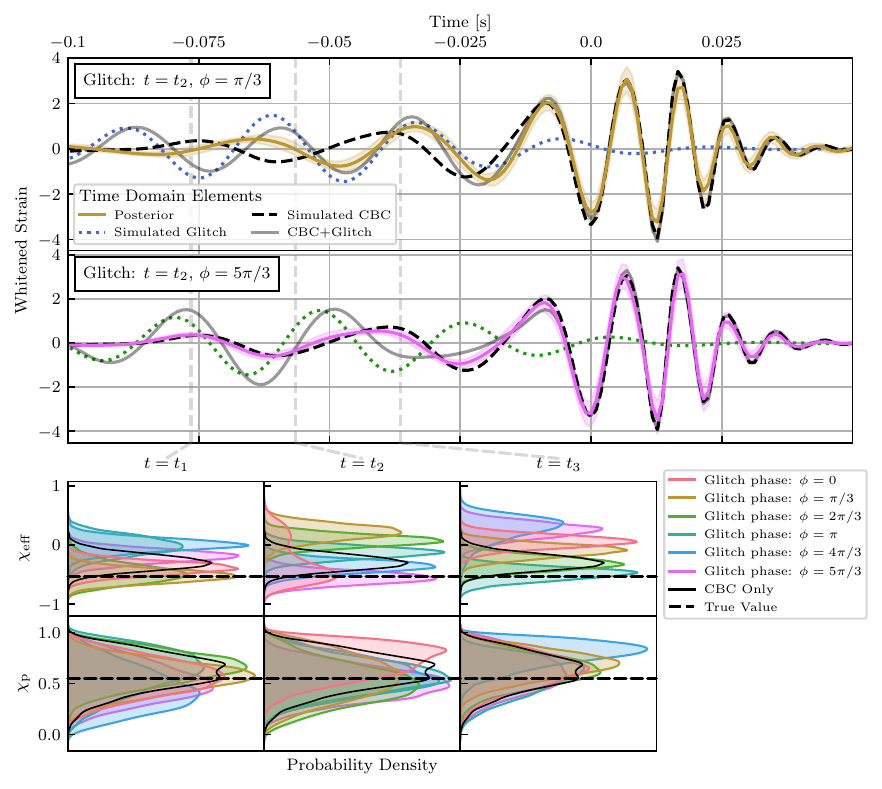}
    \caption{Same as Fig.~\ref{fig:heavy-moderate-phase-scans} for the \hmes{} configuration simulated along with glitches of varying time and phase \runsref{rid:zero-heavy-extreme-phase-start-t1}{rid:zero-heavy-extreme-phase-end-t3}.
    The highlighted cases are $t=t_2, \phi=\pi/3$ and $t=t_2, \phi=5\pi/3$.
    }\label{fig:heavy-extreme-phase-scans}
\end{figure*}

\begin{figure*}
    \centering
    \includegraphics[width=\textwidth]{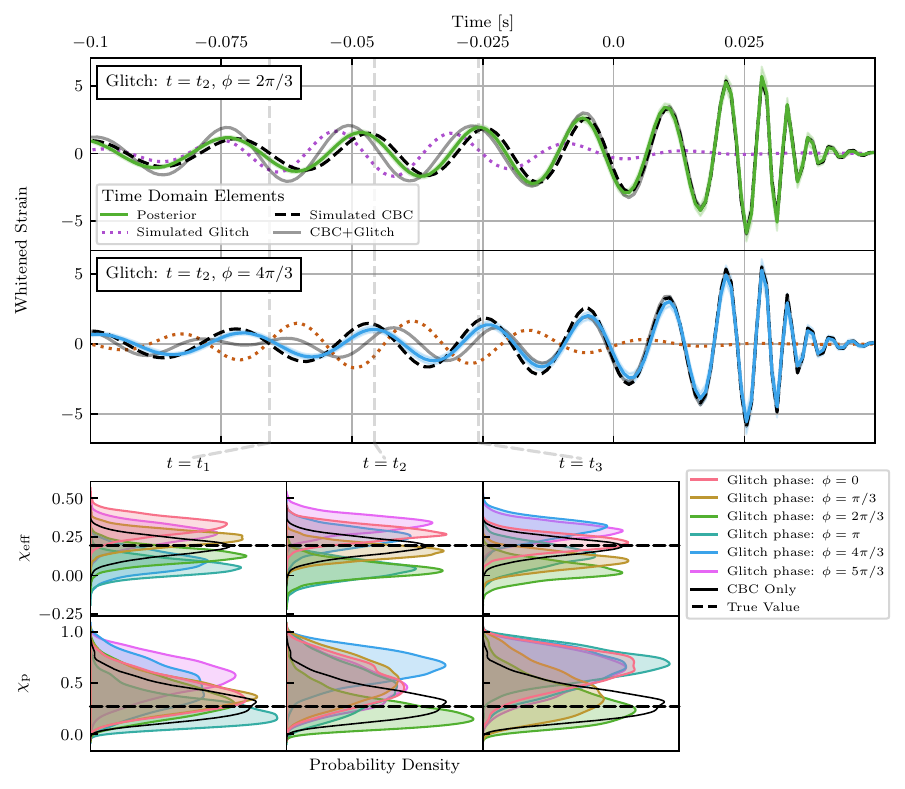}
    \caption{Same as Figure~\ref{fig:heavy-moderate-phase-scans} for a \mmms{} configuration simulated along with wavelets of varying time and phase \runsref{rid:zero-moderate-moderate-phase-start-t1}{rid:zero-moderate-moderate-phase-end-t3}.
    The highlighted cases are $t=t_2, \phi=2\pi/3$ and $t=t_2, \phi=4\pi/3$.
    }\label{fig:moderate-moderate-phase-scans}
\end{figure*}

\begin{figure*}
    \centering
    \includegraphics[width=\textwidth]{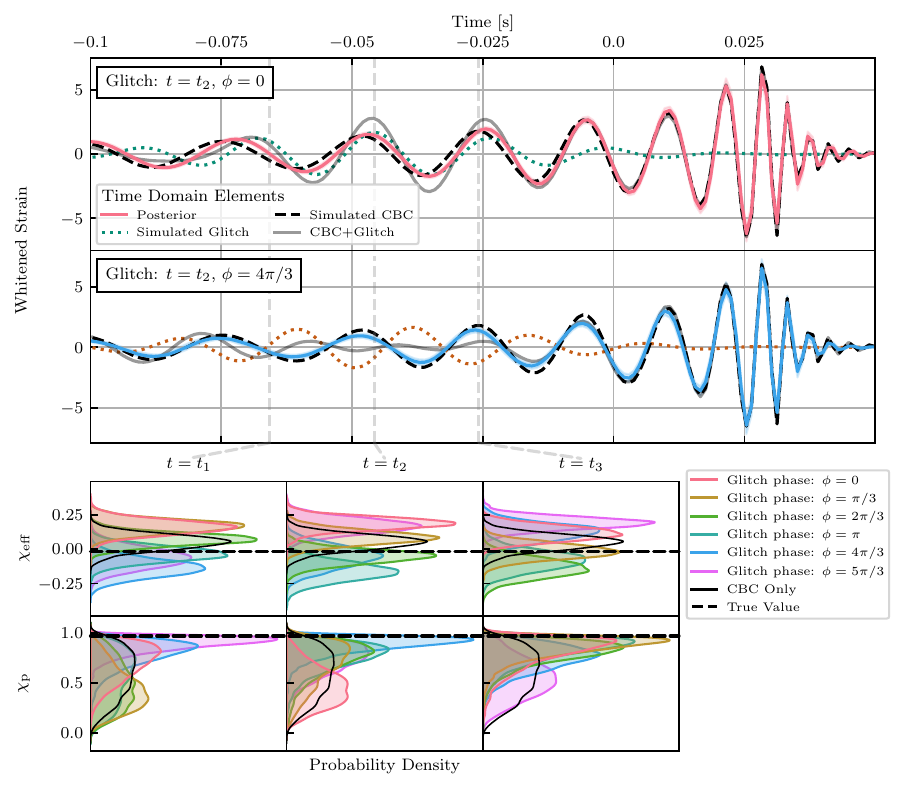}
   \caption{Same as Figure~\ref{fig:heavy-moderate-phase-scans} for a \mmes{} configuration simulated along with wavelets of varying time and phase \runsref{rid:zero-moderate-extreme-phase-start-t1}{rid:zero-moderate-extreme-phase-end-t3}.
   The highlighted cases are $t=t_2, \phi=0$ and $t=t_2, \phi=5\pi/3$.
    }\label{fig:moderate-extreme-phase-scans}
\end{figure*}

In Sec.~\ref{sec:predicting-glitch-bias} we noted that glitches cause biases in parameter inference when they sum with the true waveform to produce an aggregate that resembles a waveform for some other astrophysical configuration. 
This then leads to the question of how precisely the glitch must align with features in the CBC waveform to produce a bias, and how the biases vary as a function of this alignment.
To investigate this, we simulate glitches along with CBCs as in Sec.~\ref{sec:varying-glitch-snr}, using the same reference glitches but now fixing their SNR and instead varying their time and phase along the signal.
For each case we study three central times (labeled $t_1, t_2, t_3$ in Figs.~\ref{fig:heavy-moderate-phase-scans}---\ref{fig:moderate-extreme-phase-scans}) and six phases (evenly spaced from $0$ to $2\pi$), along with the reference case where no glitch is simulated.
We do not add Gaussian noise to isolate the effects of the glitch; for an example of this analysis in Gaussian noise, see Appendix~\ref{sec:gaussian-phase-scan}.

Figure~\ref{fig:heavy-moderate-phase-scans} shows results for the \hmms{} configuration \runsref{rid:zero-heavy-moderate-phase-start-t1}{rid:zero-heavy-moderate-phase-end-t3}, along with a glitch parameter configuration drawn from analyses of GW191109 (single wavelet with SNR $\rho_g=7.51$).
When no glitch is present, the true values of $\chieff$ and $\chip$ are recovered with \PTCHIEFF{ZHMMSNW} and \PTCHIP{ZHMMSNW} credibility respectively.
When a glitch is present, $\chieff$ cycles with the phase of the glitch from values significantly more positive than the truth to values significantly more negative.
The extrema on either end strongly exclude the true value; for example,  \PTCONLABELS{ZHMMST3P3} and \PTCONLABELS{ZHMMSMINCHIEFF} both have the true value at \PTCHIEFF{ZHMMSMIN} credibility but from the positive and negative directions respectively.
In the reconstructions, we see that for \PTCONLABELS{ZHMMST3P3} the glitch sums with the CBC to suppress a cycle approximately 0.03 seconds before the reference time, while for \PTCONLABELS{ZHMMST3P5} the glitch sums constructively with the CBC in this area.
This results in significant deviations from the simulated waveform in opposite directions for each of these, which is in turn translated to larger parameter biases. 
The pattern in $\chip$ is less cyclical, with most configurations producing similar values but a few strongly diverging from the true value, with \PTCONLABELS{ZHMMSMINCHIP} having the lowest value for the true value's credibility at \PTCHIP{ZHMMSMIN}.
A couple features are notable which will recur in the other analyses: \textit{the overlap of the glitch with the CBC in phase and time plays a large role in the magnitude and direction of biases, but there is a wide range of configurations that produce significant biases.} 

Figure~\ref{fig:heavy-extreme-phase-scans} shows results for the \hmes{} configuration \runsref{rid:zero-heavy-extreme-phase-start-t1}{rid:zero-heavy-extreme-phase-end-t3}, with the same glitch as for Fig.~\ref{fig:heavy-moderate-phase-scans}.
This configuration is characterized by a negative $\chieff$, and due to prior effects the case without a glitch prefers values nearer to 0, only finding the true value at \PTCHIEFF{ZHMESNW} credibility.
Simulations with the glitch added once again display a cyclic variation in $\chieff$, but this time distributions range from peaking at the true value---\PTCONLABELS{ZHMEST2P5} has the true value at \PTCHIEFF{ZHMEST2P5} credibility---to pushing into positive configurations, such that \PTCONLABELS{ZHMEST2P1} has the true value at \PTCHIEFF{ZHMEST2P1} credibility.
The reconstructions show a similar pattern as in Fig.~\ref{fig:heavy-moderate-phase-scans}, with the \PTCONLABELS{ZHMEST2P5} having a cycle significantly suppressed, while \PTCONLABELS{ZHMEST2P1} has that same cycle shifted and amplified.
This curious situation in which the glitch counteracts prior driven effects is similar to that in Fig.~\ref{fig:moderate-moderate-subtractions} and will also be seen in Fig.~\ref{fig:moderate-extreme-phase-scans}.
In $\chip$, the progression is very similar to that in the previous case (\hmms{}), though starting from a higher simulated value of $\chip$.
The analysis without a glitch finds the true value at \PTCHIP{ZHMESNW} credibility, while a few cases have exceptionally high $\chip$ values, with \PTCONLABELS{ZHMESMINCHIP} having the true value at \PTCHIP{ZHMESMIN} credibility.

Figure~\ref{fig:moderate-moderate-phase-scans} shows results for the \mmms{} configuration \runsref{rid:zero-moderate-moderate-phase-start-t1}{rid:zero-moderate-moderate-phase-end-t3}, with a glitch adapted to the GW200129 morphology (glitch SNR $\rho_g=7.96$).
In this configuration the CBC-only result recovers the simulated signal well, finding the true values in $\chieff$ and $\chip$ at \PTCHIEFF{ZMMMSNW} and \PTCHIP{ZHMMSNW} credibility respectively. 
Posteriors in $\chieff$ display similar cyclic behaviors to previous (high mass) cases, with the worst case being \PTCONLABELS{ZMMMSMINCHIEFF} which has the true value at \PTCHIEFF{ZHMMSMIN} credibility, though the deviation in the reconstructions is much more subtle than for the high mass case.
Posteriors in $\chip$ are again much less cyclic but have significantly larger deviations, corresponding to the choice of initial glitch which was known to affect the parameter inference for GW200129 in this manner.
The most extreme deviation is \PTCONLABELS{ZMMMSMINCHIP} which has the true value at \PTCHIP{ZMMMSMIN} credibility, but unlike the previous cases the majority of glitches at $t=t_3$ produce a similar level of bias. 

Finally, Fig.~\ref{fig:moderate-extreme-phase-scans} shows results for the \mmes{} configuration, with the same glitch used in Fig.~\ref{fig:moderate-moderate-phase-scans}.
Here, as in Fig.~\ref{fig:heavy-extreme-phase-scans}, the true configuration---in this case $\chip$---is so extreme as to be strongly disfavored by the prior, with the CBC-only analysis finding the true values in $\chieff$ and $\chip$ at \PTCHIEFF{ZMMESNW} and \PTCHIP{ZMMESNW} credibility respectively.
As in the other cases, $\chieff$ varies cyclically with the phase for a given central time, with the most extreme case being \PTCONLABELS{ZMMESMINCHIEFF} which has the true value at \PTCHIEFF{ZMMESMIN} credibility.
In $\chip$ the glitch often counteracts the effects of the prior as it did for $\chieff$ in the \hmes{} case, with most cases with a glitch injected preferring the true value with higher credibility than the CBC-only configuration, the most extreme such case being \PTCONLABELS{ZMMESMAXCHIP} which has the true value at \PTCHIP{ZMMESMAX} credibility.
However, a few cases do draw further away from the true value, such as \PTCONLABELS{ZMMESMINCHIP} which has the true value at \PTCHIP{ZMMESMIN} credibility. 

These analyses demonstrate some notable trends.
The cases we study consistently produce significant biases in the spin posteriors, typically being more severe the nearer the glitch is to the merger time, but not requiring very fine alignment of the glitch with the CBC.
While it is fairly consistent that there is a bias, the details of that bias---whether one or both of the effective spin parameters were affected, the direction of the bias, and its magnitude---depend much more sensitively on the way in which the glitch overlaps with the CBC.
In cases where spins are moderate, these biases frequently push the recovered posteriors to more extreme values.
For extreme cases these biases sometimes counteract prior effects and push the recovered posteriors nearer to the true values, while at other times moderating the recovered posterior.
Biases in $\chieff$ appear more consistently and vary cyclically with the phase of the glitch, while $\chip$ posteriors typically cluster with a few significant deviations at specific phases.
From these analyses we identified configurations which we considered interesting and used these for further study in Sec.~\ref{sec:varying-glitch-snr}.
To identify glitches for Sec.~\ref{sec:subtraction-biases} we performed the analogous analysis in the presence of Gaussian noise, the details of which can be found in Appendix~\ref{sec:gaussian-phase-scan}.

\section{Conclusions} \label{sec:conclusion}

In this paper, we presented a number of investigations into the impact of glitches on parameter inference for CBCs with an emphasis on spins, both when glitches are unmitigated and when they are mitigated with standard subtraction methods.
We demonstrated that statistical uncertainties in the glitch subtraction process will leave behind residual glitch power with non-trivial SNR, and explored how different glitch models affect those conclusions.
We showed that even at sub-detectable SNRs, glitches can produce biased spin inference.
Similarly, the output of glitch subtraction methods can \textit{also} result in biases.
Finally, we investigated how the alignment of a glitch with respect to a CBC signal (in time and phase) can alter its impacts upon parameter inference, and found that a wide variety of such alignments can produce non-trivial biases.

Two conclusions are most pressing.
Firstly, it is entirely possible --- though we cannot assess how probable without a better understanding of the glitch population --- that the inference of spins from GW events have been subtly biased by glitches of which we are not aware.
Secondly, we have shown that even for detectable glitches, the process of glitch subtraction can in some cases fail to fully mitigate biases in the inferences of spins.
This implies that spin inference downstream to glitch subtraction may not be entirely reliable, as has been previously suspected~\cite{Payne:2022spz, Udall:2024ovp}.

Of these two challenges, we are better equipped to address the second via joint CBC-glitch inference~\cite{Chatziioannou:2021ezd,Hourihane:2022doe,Udall:2024ovp}.
By contrast, new methods will likely be required to reliably resolve the danger of glitch biases \textit{sub silentio}.
Glitch marginalization may be useful here, if applied to every event consistently, though further testing would be needed to ensure that it works for glitches of low SNR where the prior on the glitch model will impact the inference~\cite{Malz:2025xdg}.
More generally, in Sec.~\ref{sec:residualSNR} it can be seen that glitches might not be recovered at all during inference when they are sufficiently quiet, so it is possible that glitch marginalization methods as currently implemented will also be insufficient.
If robust knowledge of the low-SNR glitch population is developed~\cite{Malz:2025xdg}, inference under a glitch population prior could lower the SNR threshold at which marginalization recovers a glitch.
Other potential solutions include modified likelihoods~\cite{Legin:2024gid, Ashton:2022ztk,Edy:2021par} or statistics to identify inference results that have been impacted by glitches~\cite{Udall:2024ovp}.

Through this work, we emphasized that careful consideration of glitches is \textit{always} necessary when performing and interpreting parameter inference.
We intentionally focused on worst case scenarios, using glitch configurations which are known to cause significant biases in parameter inference, and accordingly we make no statement about the probability of finding biases as severe as these in true data.
However, both glitch configurations are drawn from analyses of real events in GWTC-3, and we showed that precise fine-tuning of the glitch's phase is not necessary to produce these biases.
More work is needed to understand the population and statistics of low- and moderate-SNR glitches and to build a robust framework for mitigating their impact.

\acknowledgments

We thank Heather Fong, Mervyn Chan, Katie Rink, Sofia Alvarez, Lucy Thomas, and Evan Goetz for helpful discussions and comments.
We thank Gregory Ashton for helpful comments and suggestions during internal review. 

RU, SB, and DD were supported by NSF Grant PHY-2309200.
KC, SM, and SH were supported by NSF Grant PHY-2308770. 
SH was supported by the National Science Foundation Graduate Research Fellowship under Grant DGE-1745301. 
RU and YL were supported by the NSERC Alliance program. JM was supported by the Canada Research Chairs Program. 

This material is based upon work supported by NSF’s LIGO Laboratory 
which is a major facility fully funded by the 
National Science Foundation.
LIGO was constructed by the California Institute of Technology 
and Massachusetts Institute of Technology with funding from 
the National Science Foundation, 
and operates under cooperative agreement PHY-2309200. 
The authors are grateful
for computational resources provided by the LIGO Laboratory and supported by National Science Foundation
Grants PHY-0757058 and PHY-0823459.
 
This work made use of \textsc{Numpy}~\cite{harris2020array}, \textsc{Scipy}~\cite{2020SciPy-NMeth}, \textsc{Matplotlib}~\cite{Hunter:2007}, \textsc{Lalsuite}~\cite{lalsuite}, \textsc{Dynesty}~\cite{Speagle:2019ivv}, \textsc{Gwpy}~\cite{gwpy}, \textsc{Astropy}~\cite{2022ApJ...935..167A}, \textsc{Bilby}~\cite{bilby_paper}, \textsc{Bilby\_Pipe}~\cite{bilby_pipe_paper}, and \textsc{BayesWave}~\cite{Cornish:2014kda,Littenberg:2014oda, Cornish:2020dwh}.

\appendix

\section{The dependence of glitch bias significance on the sensitivity of the detector}\label{sec:glitch-bias-sensitivity}

Some systematic biases in measured parameters---notably biases due to waveform systematics---become more significant with improvements in detector sensitivity, as the widths of posteriors grow narrower.
Here we will show that the biases considered in this paper due to unsubtracted sub-threshold glitches or residual power after glitch subtraction do not follow this pattern.

In the linear signal regime, the shift in the best-fit parameters measurement due to the presence of a glitch and Gaussian noise is given by Eq.~\eqref{eq:parameter-error-due-to-glitch} and the covariance is 
\begin{equation}\label{eq:lsa-noise-variance}
    \langle \Delta \theta^{\alpha} \Delta \theta^\beta \rangle = (\Gamma^{-1}(\hat{\theta}^\alpha))^{\alpha \beta}\,.
\end{equation}
To understand how these quantities depend on the sensitivity of the detector, consider the uniform scaling of the noise process's amplitude by a factor $A$, such that $\mathbf{n} \rightarrow \mathbf{n}/A$.
This then results in the scalings:
\begin{eqnarray}
S_n(f) &\rightarrow \frac{S_n(f)}{A^2}\,, \\
\langle a | b \rangle &\rightarrow A^2 \langle a |  b \rangle\,, \\
\Gamma_{\alpha \beta} &\rightarrow A^2 \Gamma_{\alpha \beta}\,, \\
(\Gamma^{-1})^{\alpha \beta} &\rightarrow \frac{(\Gamma^{-1})^{\alpha \beta}}{A^2}\,.
\end{eqnarray}
\textit{Assuming that the glitch amplitude does not change}, then Eq.~\eqref{eq:parameter-error-due-to-glitch} scales as 
\begin{equation}
    \delta \theta^\alpha \propto \frac{1}{A^2} A^2 = 1\,,
\end{equation}
while Eq.~\eqref{eq:lsa-noise-variance} scales as
\begin{equation}
\langle \Delta \theta^{\alpha} \Delta \theta^\beta \rangle \propto \frac{1}{A^2}\,.
\end{equation}
The standard deviation of a given parameter measurement $\Delta \theta^\alpha$ is scaled by $\frac{1}{A}$, while the bias remains unchanged.
If one substitutes the waveform systematic error $\delta\mathbf{h}$ for $\mathbf{g}$, one sees that this is why waveform systematics become increasingly problematic as detector sensitivity increases.
However, the assumption that the glitch amplitude does not change is not actually appropriate in this case.

Post-subtraction residual SNRs and the threshold for glitch identification depend on the complexity of the model used in the subtraction and the sensitivity of the glitch detection algorithm respectively, and not the detector sensitivity.
The amplitude of the glitch or residual with some SNR will decrease along with the amplitude of the noise.
As such, scaling the glitch $\mathbf{g} \rightarrow B \mathbf{g}$, for the SNR to be constant it must be that
\begin{equation}
    \langle g |g \rangle = A^2 B^2 \langle g | g \rangle \rightarrow B = \frac{1}{A}\,.
\end{equation}
Then Eq.~\eqref{eq:parameter-error-due-to-glitch} instead scales as
\begin{equation}
     \delta \theta^\alpha \propto \frac{1}{A^2} A^2 \frac{1}{A} = \frac{1}{A}\,,
\end{equation}
which matches the scaling of the standard deviation in the parameter measurement.
The bias due to the Gaussian noise also shares this scaling.
Accordingly, the biases considered in this work will decrease in absolute terms with improving detector sensitivity, but will remain the same relative to uncertainties from Gaussian noise. 

\section{Results varying wavelet phase and time in Gaussian noise} \label{sec:gaussian-phase-scan}

\begin{figure*}
    \centering
    \includegraphics[width=\textwidth]{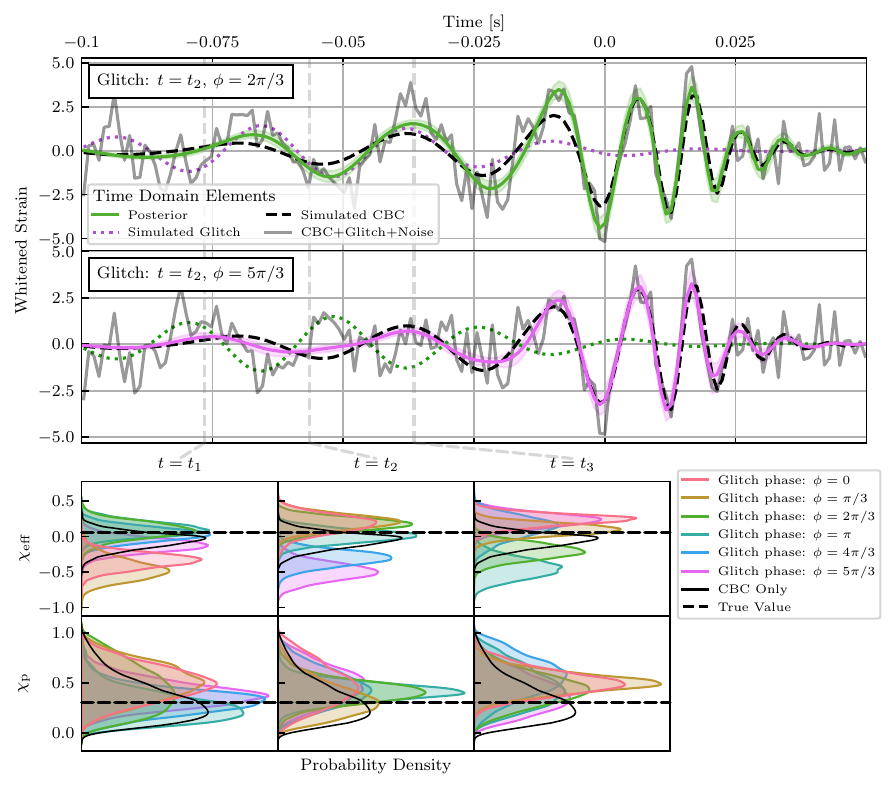}
    \caption{Same as Fig.~\ref{fig:heavy-moderate-phase-scans} for the \hmms{} configuration simulated along with glitches of varying time and phase \runsref{rid:gaussian-heavy-moderate-phase-start-t1}{rid:gaussian-heavy-moderate-phase-end-t3} in Gaussian noise.
    Here the dark gray solid line shows the sum of the data including the CBC, glitch, and Gaussian noise. 
    }\label{fig:gaussian-heavy-moderate-phase-scan}
\end{figure*}

\begin{figure*}
    \centering
    \includegraphics[width=\textwidth]{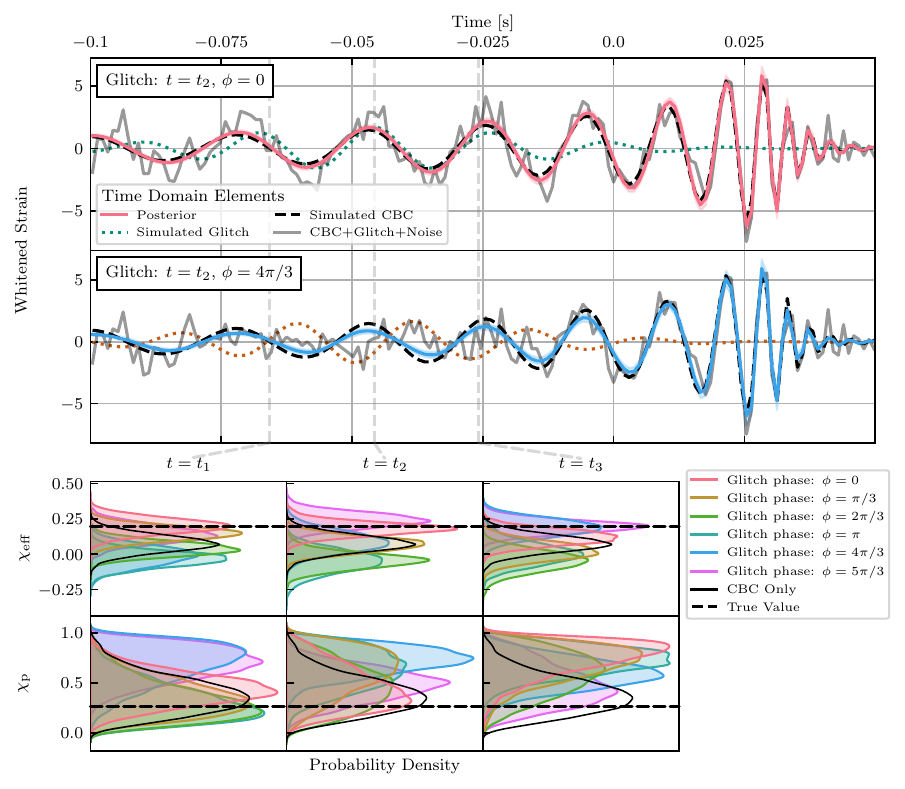}
    \caption{Same as Fig.~\ref{fig:gaussian-heavy-moderate-phase-scan} for the \mmms{} configuration simulated along with glitches of varying time and phase \runsref{rid:gaussian-moderate-moderate-phase-start-t1}{rid:gaussian-moderate-moderate-phase-end-t3} in Gaussian noise.
    }\label{fig:gaussian-moderate-moderate-phase-scan}
\end{figure*}

In Sec.~\ref{sec:subtraction-biases} we studied analyses in Gaussian noise, in order to emulate glitch subtraction in real data.
The interaction of glitches with parameter inference is modulated by the inclusion of Gaussian noise, both in the overall trends and in the ordering of which configurations result in which trends.
Accordingly, to identify the best examples to explore further in Sec.~\ref{sec:subtraction-biases} we repeat the analysis of Sec.~\ref{sec:phasetime} using Gaussian data.

Figure~\ref{fig:gaussian-heavy-moderate-phase-scan} shows the results for the \hmms{} configuration simulated with glitches in Gaussian noise \runsref{rid:gaussian-heavy-moderate-phase-start-t1}{rid:gaussian-heavy-moderate-phase-end-t3}.
By and large these results are analogous to those in Fig.~\ref{fig:heavy-moderate-phase-scans}.
In $\chieff$, the magnitudes of some shifts are changed from the zero noise results at a small level, including for the reference case \runref{rid:gaussian-heavy-moderate} which found the true value at \PTCHIEFF{GHMMSNW} credibility, but configurations consistently reproduce the biases they experienced in the zero noise configuration.
Results in $\chip$ vary significantly more, primarily by pulling all of the posteriors towards more moderate values.
Some posteriors are strongly sharpened, for example \PTCONLABELS{GHMMST3P1}, which fully excludes $\chip=0$ and has the true value at \PTCHIP{GHMMST3P1} credibility.
For the purposes of Sec.~\ref{sec:subtraction-biases} we choose to emphasize biases in $\chieff$, and hence select the \PTCONLABELS{GHMMST3P3} case---which has the true value of $\chieff$ at \PTCHIEFF{GHMMST3P3} credibility---for further analysis.

Figure~\ref{fig:gaussian-moderate-moderate-phase-scan} shows results for the \mmms{} configuration simulated with glitches in Gaussian noise \runsref{rid:gaussian-moderate-moderate-phase-start-t1}{rid:gaussian-moderate-moderate-phase-end-t3}.
In this case, as noted in Sec.~\ref{sec:subtraction-biases}, the reference case \runref{rid:gaussian-moderate-moderate} behaves oddly, with $\chieff$ posteriors shifted by Gaussian noise to be below the true value, thus finding it at only \PTCHIEFF{GMMMSNW} credibility.
This propagates to all glitch-affected posteriors, which maintain similar profiles and spreads, but are shifted downwards with respect to the true posteriors. 
The most extreme case, \PTCONLABELS{GMMMSMINCHIEFF}, still fully excludes the true value, having it at \PTCHIEFF{GMMMSMIN} credibility.
Meanwhile in $\chip$ the CBC-only case does recover the true value with \PTCHIP{GMMMSNW} credibility, but there is greater variety in the glitch posteriors, with the most extreme being \PTCONLABELS{GMMMSMINCHIP} which has the true value at \PTCHIP{GMMMSMIN} credibility.
For the purposes of Sec.~\ref{sec:subtraction-biases} we choose to emphasize biases in $\chip$, and hence select the \PTCONLABELS{GMMMST3P0} case---which has the true value of $\chip$ credibility---for further analysis.

\section{Detailed analysis settings}
\label{app:settings}

In this section we present a tabulation of runs which were performed in this paper. 
Analyses are categorized by the properties of any glitch which was simulated (where applicable), the properties of the glitch inference method (where applicable), and the configuration of the CBC model used (where applicable).
Whenever a CBC was simulated, it was also inferred by the same model with which it was injected. 
Data for these analyses are made public in the associated data-set~\cite{dataset}.

\begin{longtable*}{|c|>{\centering\arraybackslash}p{15mm}|>{\centering\arraybackslash}p{18mm}|>{\centering\arraybackslash}p{16mm}|>{\centering\arraybackslash}p{28mm}|>{\centering\arraybackslash}p{22mm}|>{\centering\arraybackslash}p{28mm}|>{\centering\arraybackslash}p{16mm}|}
\hline
\textbf{Run ID} & \textbf{Noise Type} & \textbf{Simulated Glitch Model} & \textbf{Simulated Glitch SNR} & \textbf{Simulated Glitch Parameters} & \textbf{Glitch Inference Model} & \textbf{CBC Configuration} & \textbf{Relevant Figures}\\\hhline{|=|=|=|=|=|=|=|=|}\endhead

\ridref{rid:slow-scattering-snr-5}\addtocounter{RunIDCounter}{4}-\ridref{rid:slow-scattering-snr-50} & Gaussian & Slow scattering & 5, 10, 20, 30, 40, 50 & for each SNR, 50 drawn from priors & Slow scattering & --- & \ref{fig:slow-scattering-residual-snr}\\ \hline

\ridref{rid:BW_1-start}\addtocounter{RunIDCounter}{3}-\ridref{rid:BW_1-end} & Gaussian & Wavelet $D_{\mathrm{inj}}=1$ & 5, 10, 20, 30, 40 & for each SNR, 50 drawn from priors & Wavelet $D_{\mathrm{rec}}=1$ & --- &\ref{fig:BW_residual}\\ \hline
\ridref{rid:BW_2}\addtocounter{RunIDCounter}{3}-\ridref{rid:BW_2-end} & Gaussian & Wavelet $D_{\mathrm{inj}}=3$ & 5, 10, 20, 30, 40 & for each SNR, 50 drawn from priors & Wavelet $D_{\mathrm{rec}}=3$ & --- &\ref{fig:BW_residual}\\ \hline
\ridref{rid:BW_3}\addtocounter{RunIDCounter}{3}-\ridref{rid:BW_3-end} & Gaussian & Wavelet $D_{\mathrm{inj}}=10$ &  5, 10, 20, 30, 40  & for each SNR, 50 drawn from priors & Wavelet $D_{\mathrm{rec}}=10$ & --- & \ref{fig:BW_residual}\\ \hline
\ridref{rid:BW_4}\addtocounter{RunIDCounter}{3}-\ridref{rid:BW_4-end} & Gaussian & Wavelet $D_{\mathrm{inj}}=10$ & 5, 10, 20, 30, 40  & for each SNR, 50 drawn from priors & Wavelet $D_{\mathrm{rec}}=10$ & --- & \ref{fig:BW_residual} \\ \hline
\ridref{rid:BW_5}\addtocounter{RunIDCounter}{3}-\ridref{rid:BW_5-end} & Gaussian & Wavelet $D_{\mathrm{inj}}=10$ & 5, 10, 20, 30, 40 & for each SNR, 50 drawn from priors & Wavelet $D_{\mathrm{rec}}=(0,10)$ & --- & \ref{fig:BW_residual} \\ \hline
\ridref{rid:BW_6}\addtocounter{RunIDCounter}{3}-\ridref{rid:BW_6-end} & Gaussian & Wavelet $D_{\mathrm{inj}}=1$ &  5, 10, 20, 30, 40  & for each SNR, 50 drawn from priors & Wavelet $D_{\mathrm{rec}}=(0,5)$ & --- &\ref{fig:BW_residual} \\ \hline
\ridref{rid:BW_7}\addtocounter{RunIDCounter}{3}-\ridref{rid:BW_7-end} & Gaussian & Wavelet $D_{\mathrm{inj}}=3$ &  5, 10, 20, 30, 40  & for each SNR, 50 drawn from priors & Wavelet $D_{\mathrm{rec}}=(0,5)$ & --- & \ref{fig:BW_residual} \\ \hline

\ridref{rid:zero-heavy-moderate} & Zero & --- & --- & --- & --- & \hmms &\ref{fig:heavy-moderate-varying-snr}, \ref{fig:heavy-moderate-phase-scans}\\ \hline
\ridref{rid:zero-heavy-extreme} & Zero & --- & --- & --- & --- & \hmes & \ref{fig:heavy-extreme-varying-snr}, \ref{fig:heavy-extreme-phase-scans}\\ \hline
\ridref{rid:zero-moderate-moderate} & Zero & --- & --- & --- & --- & \mmms & \ref{fig:moderate-moderate-varying-snr}, \ref{fig:moderate-moderate-phase-scans}\\ \hline
\ridref{rid:zero-moderate-extreme} & Zero & --- & --- & --- & --- & \mmes & \ref{fig:moderate-extreme-varying-snr}, \ref{fig:moderate-extreme-phase-scans}\\ \hline

\ridref{rid:heavy-moderate-snr-start}\addtocounter{RunIDCounter}{8}-\ridref{rid:heavy-moderate-snr-end} & Zero & Wavelet & 1, 2, 3, 4, 5, 6, 7, 8, 9, 10 & $f_0=33.6$ Hz, $Q=8.74$,$\phi=\frac{5\pi}{3}$, $t=t_2$ & --- & \hmms & \ref{fig:heavy-moderate-varying-snr}\\ \hline

\ridref{rid:heavy-extreme-snr-start}\addtocounter{RunIDCounter}{8}-\ridref{rid:heavy-extreme-snr-end} & Zero & Wavelet & 1, 2, 3, 4, 5, 6, 7, 8, 9, 10 & $f_0=33.6$ Hz, $Q=8.74$, $\phi=\frac{\pi}{3}$, $t=t_3$ & --- & \hmes & \ref{fig:heavy-extreme-varying-snr}\\ \hline

\ridref{rid:moderate-moderate-snr-start}\addtocounter{RunIDCounter}{8}-\ridref{rid:moderate-moderate-snr-end} & Zero & Wavelet & 1, 2, 3, 4, 5, 6, 7, 8, 9, 10 & $f_0=43.6$ Hz, $Q=10.7$, $\phi=\pi$, $t=t_3$ & --- & \mmms & \ref{fig:moderate-moderate-varying-snr}\\ \hline

\ridref{rid:moderate-extreme-snr-start}\addtocounter{RunIDCounter}{8}-\ridref{rid:moderate-extreme-snr-end} & Zero & Wavelet & 1, 2, 3, 4, 5, 6, 7, 8, 9, 10 & $f_0=43.6$ Hz, $Q=10.7$, $\phi=0$, $t=t_2$ & --- & \mmes & \ref{fig:moderate-extreme-varying-snr} \\ \hline

\ridref{rid:gaussian-heavy-moderate} & Gaussian & --- & --- & --- & --- & \hmms & ~\ref{fig:heavy-moderate-subtractions}~\ref{fig:gaussian-heavy-moderate-phase-scan}\\ \hline
\ridref{rid:gaussian-heavy-moderate-phase-start-t1}\addtocounter{RunIDCounter}{16}-\ridref{rid:gaussian-heavy-moderate-phase-end-t3} & Gaussian & Wavelet & 7.51 & $f_0=33.6$ Hz, $Q=8.74$, $\phi\in\{\frac{k\pi}{3}, k=0$\textemdash$5\}$ $\times \;t\in\{t_1, t_2, t_3\}$ & --- & \hmms & ~\ref{fig:heavy-moderate-subtractions}~\ref{fig:gaussian-heavy-moderate-phase-scan}\\ \hline
\ridref{rid:heavy-moderate-wavelet-inference} & Gaussian & Wavelet & 7.51 & $\phi=\pi$, $t=t_3$, $f_0=33.6$ Hz, $Q=8.74$ & Wavelet $D_\mathrm{rec}=1$ & --- & ~\ref{fig:heavy-moderate-subtractions}\\ \hline
\ridref{rid:heavy-moderate-maxL-subtraction} & Gaussian & Maximum likelihood residual from run~\ref{rid:heavy-moderate-wavelet-inference}& 4.57 & --- & --- & \hmms & \ref{fig:heavy-moderate-subtractions} \\ \hline
\ridref{rid:heavy-moderate-median-subtraction} & Gaussian & Pointwise median residual from run~\ref{rid:heavy-moderate-wavelet-inference}& 4.27 & --- & --- & \hmms & \ref{fig:heavy-moderate-subtractions} \\ \hline
\ridref{rid:heavy-moderate-fairdraw1-subtraction} & Gaussian & Fairdraw 1 residual from run~\ref{rid:heavy-moderate-wavelet-inference}& 4.65 & --- & --- & \mmms & \ref{fig:heavy-moderate-subtractions} \\ \hline
\ridref{rid:heavy-moderate-fairdraw2-subtraction} & Gaussian & Fairdraw 2 residual from run~\ref{rid:heavy-moderate-wavelet-inference}& 4.29 & --- & --- & \hmms & \ref{fig:heavy-moderate-subtractions} \\ \hline
\ridref{rid:heavy-moderate-fairdraw3-subtraction} & Gaussian & Fairdraw 3 residual from run~\ref{rid:heavy-moderate-wavelet-inference}& 5.95 & --- & --- & \hmms & \ref{fig:heavy-moderate-subtractions} \\ \hline
\ridref{rid:heavy-moderate-fairdraw4-subtraction} & Gaussian & Fairdraw 4 residual from run~\ref{rid:heavy-moderate-wavelet-inference}& 6.29 & --- & --- & \hmms & \ref{fig:heavy-moderate-subtractions} \\ \hline
\ridref{rid:heavy-moderate-fairdraw5-subtraction} & Gaussian & Fairdraw 5 residual from run~\ref{rid:heavy-moderate-wavelet-inference}& 4.58 & --- & --- & \hmms & \ref{fig:heavy-moderate-subtractions} \\ \hline
\ridref{rid:heavy-moderate-marginalized} & Gaussian & Wavelet & 7.51 & $\phi=\pi$, $t=t_3$, $f_0=33.6$ Hz, $Q=8.74$ & Wavelet $D_\mathrm{rec}=1$ & \hmms & ~\ref{fig:heavy-moderate-subtractions}\\ \hline
\ridref{rid:heavy-moderate-bayeswave-marginalized} & Gaussian & Wavelet & 7.51 & $\phi=\pi$, $t=t_3$, $f_0=33.6$ Hz, $Q=8.74$ & Wavelet $D_\mathrm{rec}=(0, 10)$ & \hmms & ~\ref{fig:heavy-moderate-subtractions}\\ \hline

\ridref{rid:gaussian-moderate-moderate} & Gaussian & --- & --- & --- & --- & \mmms &~\ref{fig:moderate-moderate-subtractions}~\ref{fig:gaussian-moderate-moderate-phase-scan}\\ \hline
\ridref{rid:gaussian-moderate-moderate-phase-start-t1}\addtocounter{RunIDCounter}{16}-\ridref{rid:gaussian-moderate-moderate-phase-end-t3} & Gaussian & Wavelet & 7.96 & $f_0=43.6$ Hz, $Q=10.7$, $\phi\in\{\frac{k\pi}{3}, k=0$\textemdash$5\}$ $\times \;t\in\{t_1, t_2, t_3\}$ & --- & \mmms & ~\ref{fig:moderate-moderate-subtractions}~\ref{fig:gaussian-moderate-moderate-phase-scan}\\ \hline
\ridref{rid:moderate-moderate-wavelet-inference} & Gaussian & Wavelet & 7.96 & $\phi=0$, $t=t_3$, $f_0=43.6$ Hz, $Q=10.7$& Wavelet $D_\mathrm{rec}=1$ & --- & ~\ref{fig:moderate-moderate-subtractions}\\ \hline
\ridref{rid:moderate-moderate-maxL-subtraction} & Gaussian & Maximum likelihood residual from run~\ref{rid:moderate-moderate-wavelet-inference}& 3.50 & --- & --- & \mmms & \ref{fig:moderate-moderate-subtractions} \\ \hline
\ridref{rid:moderate-moderate-median-subtraction} & Gaussian & Pointwise median residual from run~\ref{rid:moderate-moderate-wavelet-inference}& 3.37 & --- & --- & \mmms & \ref{fig:moderate-moderate-subtractions} \\ \hline
\ridref{rid:moderate-moderate-fairdraw1-subtraction} & Gaussian & Fairdraw 1 residual from run~\ref{rid:moderate-moderate-wavelet-inference}& 3.85 & --- & --- & \mmms & \ref{fig:moderate-moderate-subtractions} \\ \hline
\ridref{rid:moderate-moderate-fairdraw2-subtraction} & Gaussian & Fairdraw 2 residual from run~\ref{rid:moderate-moderate-wavelet-inference}& 5.67 & --- & --- & \mmms & \ref{fig:moderate-moderate-subtractions} \\ \hline
\ridref{rid:moderate-moderate-fairdraw3-subtraction} & Gaussian & Fairdraw 3 residual from run~\ref{rid:moderate-moderate-wavelet-inference}& 4.45 & --- & --- & \mmms & \ref{fig:moderate-moderate-subtractions} \\ \hline
\ridref{rid:moderate-moderate-fairdraw4-subtraction} & Gaussian & Fairdraw 4 residual from run~\ref{rid:moderate-moderate-wavelet-inference}& 3.49 & --- & --- & \mmms & \ref{fig:moderate-moderate-subtractions} \\ \hline
\ridref{rid:moderate-moderate-fairdraw5-subtraction} & Gaussian & Fairdraw 5 residual from run~\ref{rid:moderate-moderate-wavelet-inference}& 3.54 & --- & --- & \mmms & \ref{fig:moderate-moderate-subtractions} \\ \hline
\ridref{rid:moderate-moderate-marginalized} & Gaussian & Wavelet & 7.96 & $\phi=0$, $t=t_3$, $f_0=43.6$ Hz, $Q=10.7$& Wavelet $D_\mathrm{rec}=1$ & \mmms & ~\ref{fig:moderate-moderate-subtractions}\\ \hline
\ridref{rid:moderate-moderate-bayeswave-marginalized} & Gaussian & Wavelet & 7.96 & $\phi=0$, $t=t_3$, $f_0=43.6$ Hz, $Q=10.7$& Wavelet $D_\mathrm{rec}=(0, 10)$ & \mmms & ~\ref{fig:moderate-moderate-subtractions}\\ \hline

\ridref{rid:zero-heavy-moderate-phase-start-t1}\addtocounter{RunIDCounter}{16}-\ridref{rid:zero-heavy-moderate-phase-end-t3} & Zero & Wavelet & 7.51 & $f_0=33.6$ Hz, $Q=8.74$, $\phi\in\{\frac{k\pi}{3}, k=0$\textemdash$5\}$ $\times \;t\in\{t_1, t_2, t_3\}$ & --- & \hmms & \ref{fig:heavy-moderate-phase-scans}\\ \hline
\ridref{rid:zero-heavy-extreme-phase-start-t1}\addtocounter{RunIDCounter}{16}-\ridref{rid:zero-heavy-extreme-phase-end-t3} & Zero & Wavelet & 7.51 & $f_0=33.6$ Hz, $Q=8.74$, $\phi\in\{\frac{k\pi}{3}, k=0$\textemdash$5\}$ $\times \;t\in\{t_1, t_2, t_3\}$ & --- & \hmes & \ref{fig:heavy-extreme-phase-scans}\\ \hline
\ridref{rid:zero-moderate-moderate-phase-start-t1}\addtocounter{RunIDCounter}{16}-\ridref{rid:zero-moderate-moderate-phase-end-t3} & Zero & Wavelet & 7.96 & $f_0=43.6$ Hz, $Q=10.7$, $\phi\in\{\frac{k\pi}{3}, k=0$\textemdash$5\}$ $\times \;t\in\{t_1, t_2, t_3\}$ & --- & \mmms & \ref{fig:moderate-moderate-phase-scans}\\ \hline
\ridref{rid:zero-moderate-extreme-phase-start-t1}\addtocounter{RunIDCounter}{16}-\ridref{rid:zero-moderate-extreme-phase-end-t3} & Zero & Wavelet & 7.96 &$f_0=43.6$ Hz, $Q=10.7$, $\phi\in\{\frac{k\pi}{3}, k=0$\textemdash$5\}$ $\times \;t\in\{t_1, t_2, t_3\}$ & --- & \mmes & \ref{fig:moderate-extreme-phase-scans} \\ \hline

\caption[width=\linewidth]{A tabulation of all runs performed for this paper, designated by their run IDs.
In the cases of IDs~\ref{rid:slow-scattering-snr-5}---~\ref{rid:BW_7-end}, each run ID corresponds to 50 identically constructed analyses, in which a glitch was simulated using parameters drawn from their respective priors.
The noise type column describes whether there was underlying Gaussian noise added (Gaussian) or not (Zero).
Whenever a CBC is simulated---such that the CBC configuration column is not empty---the CBC parameters are also inferred.
}\label{tab:table-of-runs}
\end{longtable*}

\bibliography{references}

\end{document}